\documentclass[lettersize,journal]{IEEEtran}

\def\review{1} % With 1 comments are enabled, with 0 not

\if\review1
\usepackage{todonotes}
\else
\usepackage[disable]{todonotes}
\fi

\usepackage{amsmath}% for double integral symbol in this template
\usepackage{xcolor}
\usepackage{times}% use times font for the paper instead of default LaTeX fonts
\usepackage{graphicx}% for figures
\usepackage{multirow}% to allow multiple-row elements in tabular environment
\usepackage{siunitx}[mode=match]
\usepackage{hhline}
\usepackage{hyperref}
\usepackage{amsmath}
\usepackage{booktabs}
\usepackage[none]{hyphenat}% turn off hyphenation to make text extraction and indexing easier
\usepackage{float}% better control of floating figures and tables
\usepackage{subfig}% for labelling subfigures within figures
\usepackage{iftex}% for detecting whether XeLaTeX or pdfLaTeX is being used, for the purpose of choosing the right fonts below.
\usepackage{amsmath,amsfonts}
\usepackage{algorithmic}
\usepackage{algorithm}
\usepackage{array}
\usepackage{placeins}
\usepackage{cite}
\usepackage{amssymb,mathtools}
\usepackage{algorithm}
\usepackage{pgfplots} %latex plots
\usepackage{circuitikz}[european] %Circuit Library
\usepackage[percent]{overpic}
\usepackage{graphicx}
\usepackage{amsmath}
\usepackage{psfrag}
\usepackage{subfig}
\usepackage{stfloats}
\usepackage{acronym}
\usepackage[font=footnotesize]{caption}
\usepackage{xcolor}     % For colored text/legend (optional but recommended)
\usepackage{subcaption} % For side-by-side figures
\usepackage{trfsigns} % Correct formatting of e and j
\usepackage{url}

\newacro{3GPP}{3rd Generation Partnership Project}
\newacro{5G}{fifth generation}
\newacro{5G NR}{Fifth Generation New Radio}
\newacro{6G}{Sixth Generation}
\newacro{A/D}{analog-to-digital}
\newacro{AAL}{array aperture line}
\newacro{ABE}{analog back-end}
\newacro{ADC}{analog-to-digital converter}
\newacro{ADS}{Advanced Design System} 
\newacro{AFE}{analog front-end}
\newacro{AGV}{automatic guided vehicle}
\newacro{AM-AM}{amplitude-to-amplitude modulation}
\newacro{AM-PM}{amplitude-to-phase modulation}
\newacro{AWGN}{additive white Gaussian noise}
\newacro{B5G}{beyond \ac{5G}}
\newacro{BB}{baseband}
\newacro{BER}{bit error ratio}
\newacro{BPSK}{binary phase-shift keying}
\newacro{BP}{band-pass}
\newacro{BS}{base station}
\newacro{CC}{coherent compensation}
\newacro{CDM}{code-division multiplexing}
\newacro{CFO}{carrier frequency offset}
\newacro{CFR}{channel frequency response}
\newacro{CFAR}{constant false alarm rate}
\newacro{CIR}{channel impulse response}
\newacro{CoMP}{coordinated multipoint}
\newacro{CP}{cyclic prefix}
\newacro{CPE}{common phase error}
\newacro{CPO}{carrier phase offset}
\newacro{CRLB}{Cram\'er-Rao lower bound}
\newacro{CS}{chirp sequence}
\newacro{CSI}{channel state information}
\newacro{CW}{continuous wave}
\newacro{CZT}{chirp Z-transform}
\newacro{D/A}{digital-to-analog}
\newacro{DAC}{digital-to-analog converter}
\newacro{DC}{direct current}
\newacro{DDC}{digital down-conversion}
\newacro{DDS}{direct digital synthesis}
\newacro{DFRC}{dual-function radar-communication or dual-functional radar-communication}
\newacro{DFnT}{discrete Fresnel transform}
\newacro{DFT}{discrete Fourier transform}
\newacro{DL}{downlink}
\newacro{DMRS}{demodulation reference signal}
\newacro{DoA}{direction-of-arrival}
\newacro{DoD}{direction-of-departure}
\newacro{DPD}{digital pre-distortion}
\newacro{DUC}{digital up-conversion}
\newacro{ETSI}{European Telecommunications Standards Institute}
\newacro{EVM}{error vector magnitude}
\newacro{FDE}{frequency-domain equalization}
\newacro{FDM}{frequency-division multiplexing}
\newacro{FDCC}{frequency domain coherent compensation}
\newacro{FFT}{fast Fourier transform}
\newacro{FIR}{finite impulse response}
\newacro{FR-SW}{full reconstruction-based sliding window}
\newacro{FO}{frequency offset}
\newacro{FR2}{Frequency Range 2}
\newacro{gNB}{gNodeB}
\newacro{HP}{high-pass}
\newacro{HP-SIC-FDCC}{high-precision self interference cancellation frequency domain coherent compensation}
\newacro{IBFD}{in-band full duplex}
\newacro{ICI}{intercarrier interference}
\newacro{IDFT}{inverse discrete Fourier transform}
\newacro{IFFT}{inverse fast Fourier transform}
\newacro{IDFnT}{inverse discrete Fresnel transform}
\newacro{IF}{intermediate frequency}
\newacro{IHE}{Institute of Radio Frequency Engineering and Electronics}
\newacro{I/Q}{in-phase/quadrature}
\newacro{I}{in-phase}
\newacro{Q}{quadrature}
\newacro{IBO}{input back-off}
\newacro{IP1dB}{input-referred 1-dB compression point}
\newacro{ISAC}{integrated sensing and communication}
\newacro{ISI}{intersymbol interference}
\newacro{ISLR}{integrated-sidelobe level ratio}
\newacro{IM3}{third-order intermodulation}
\newacro{IoT}{Internet of Things}
\newacro{JCAS}{joint communication and sensing}
\newacro{KIT}{Karlsruhe Institute of Technology}
\newacro{KPI}{key performance indicator}
\newacro{LDPC}{low-density parity-check}
\newacro{LFSR}{linear-feedback shift register}
\newacro{LNA}{low-noise amplifier}
\newacro{LTE}{long term evolution}
\newacro{LO}{local oscillator}
\newacro{LoS}{line-of-sight}
\newacro{LP}{low-pass}
\newacro{LPF}{low-pass filter}
\newacro{LPI}{low probability of intercept}
\newacro{LS}{least squares}
\newacro{MAE}{mean absolute error}
\newacro{mmWave}{milimeter wave}
\newacro{MIMO}{multiple-input multiple-output}
\newacro{MLE}{maximum likelihood estimator}
\newacro{MLS}{maximum-length sequence}
\newacro{MRC}{maximal-ratio combining}
\newacro{MUSIC}{multiple signal classification}
\newacro{MTCC}{multi target coherent compensation}
\newacro{IMD}{intermodulation distortion}
\newacro{JIC-CC}{joint-interference cancellation with coherent compensation}
\newacro{NAF}{normalized angular frequency}
\newacro{NB}{narrowband}
\newacro{NLoS}{non-line-of-sight}
\newacro{NR}{new radio}
\newacro{OCDM}{orthogonal chirp-division multiplexing}
\newacro{OFDM}{orthogonal frequency-division multiplexing}
\newacro{OOB}{out-of-band}
\newacro{OTA}{over-the-air}
\newacro{P/S}{parallel-to-serial}
\newacro{PA}{power amplifier}
\newacro{PACF}{periodic autocorrelation function}
\newacro{PAPR}{peak-to-average power ratio}
\newacro{PCCF}{periodic cross-correlation function}
\newacro{PLC}{powerline communication}
\newacro{PLL}{phase-locked loop}
\newacro{PMCW}{phase-modulated continuous wave}
\newacro{PMN}{perceptive mobile network}
\newacro{PN}{oscillator phase noise}
\newacro{PoC}{proof-of-concept}
\newacro{PPLR}{peak power loss ratio}
\newacro{PRBS}{pseudorandom binary sequence}
\newacro{PRS}{positioning reference signal}
\newacro{PSD}{power spectral density}
\newacro{PSF}{point spread function}
\newacro{PSLR}{peak-to-sidelobe level ratio}
\newacro{QPSK}{quadrature phase-shift keying}
\newacro{RAN}[RAN]{Radio Access Network}
\newacro{ROI}{region of interest}
\newacro{RadCom}{radar-communication}
\newacro{RCS}{radar cross section}
\newacro{RF}{radio-frequency}
\newacro{RFS}{random finite set}
\newacro{RIS}{reflective intelligent surface}
\newacro{RMS}{root mean square}
\newacro{RMSE}{root mean squared error}
\newacro{RX}{receiver}
\newacro{SC}[S\&C]{Schmidl \& Cox}
\newacro{SFO}{sampling frequency offset}
\newacro{SIC}{successive-interference cancellation}
\newacro{SINR}{signal-to-interference-plus-noise ratio}
\newacro{SIR}{signal-to-interference ratio}
\newacro{SISO}{single-input single-output}
\newacro{SW}{sliding window}
\newacro{SJ}{sampling jitter}
\newacro{SNR}{signal-to-noise ratio}
\newacro{SoC}{system-on-a-chip}
\newacro{SQNR}{signal-to-quantization-noise ratio}
\newacro{SSB}{synchronization signal block}
\newacro{STO}{symbol time offset}
\newacro{S/P}{serial-to-parallel}
\newacro{TDCC}{time domain coherent compensation}
\newacro{TDD}{time-division duplexing}
\newacro{TDE}{time-domain equalization}
\newacro{TDM}{time-division multiplexing}
\newacro{TDR}{time-domain reflectometry}
\newacro{TITO}{tilt inference of time offset}
\newacro{TO}{time offset}
\newacro{TR}{technical report}
\newacro{TS}{technical specification}
\newacro{TX}{transmitter}
\newacro{UE}{user equipment}
\newacro{UL}{uplink}
\newacro{ULA}{uniform linear array}
\newacro{V2V}{vehicle-to-vehicle}
\newacro{ZF}{zero forcing}
\newacro{ZP}{zero padding}

\usepackage{fancyhdr}
\fancypagestyle{empty}{fancy}

\begin{document}

\title{Breaking the CP Limit: Robust Long-Range\\ OFDM Sensing via Interference Cleaning}

\author{Umut~Utku~Erdem,~\IEEEmembership{Graduate~Student~Member,~IEEE},
        Lucas~Giroto,~\IEEEmembership{Member,~IEEE},
        Benedikt~Geiger,~\IEEEmembership{Graduate~Student~Member,~IEEE},        Taewon~Jeong,~\IEEEmembership{Graduate~Student~Member,~IEEE},
        
        Silvio~Mandelli,~\IEEEmembership{Member,~IEEE},  
        Christian~Karle,~\IEEEmembership{Graduate~Student~Member,~IEEE}, 
        Benjamin~Nuss,~\IEEEmembership{Senior~Member,~IEEE}, Laurent~Schmalen,~\IEEEmembership{Fellow,~IEEE},
        and~Thomas~Zwick,~\IEEEmembership{Fellow,~IEEE}

        \thanks{The authors acknowledge the financial support by the Federal Ministry of Research, Technology and Space of Germany in the projects ``KOMSENS-6G'' (grant number: 16KISK123), ``SENSATION'' (grant number: 16KIS2528)  and ``Open6GHub'' (grant number: 16KISK010). \textit{(Corresponding author: Umut Utku Erdem.)}}
		\thanks{U. U. Erdem, T. Jeong, and T. Zwick are with the Institute of Radio Frequency Engineering and Electronics (IHE), Karlsruhe Institute of Technology (KIT), 76131 Karlsruhe, Germany (e-mail: {umut.erdem@kit.edu}, {taewon.jeong@kit.edu}, {thomas.zwick@kit.edu}).}
        \thanks{L. Giroto was with the Institute of Radio Frequency Engineering and Electronics (IHE), Karlsruhe Institute of Technology (KIT), 76131 Karlsruhe, Germany. He is now with Nokia Bell Labs, 70469 Stuttgart, Germany (e-mail: {lucas.giroto@nokia-bell-labs.com}).}
		\thanks{B. Geiger and L. Schmalen are with the Communications Engineering Laboratory (CEL), Karlsruhe Institute of Technology (KIT), 76187 Karlsruhe, Germany (e-mail: {benedikt.geiger@kit.edu}, {laurent.schmalen@kit.edu}).}
        \thanks{S. Mandelli is with Nokia Bell Labs, 70469 Stuttgart, Germany (e-mail: {silvio.mandelli@nokia-bell-labs.com}).}
		\thanks{C. Karle is with the Institute for Information Processing Technology (ITIV), Karlsruhe Institute of Technology (KIT), 76131 Karlsruhe, Germany (e-mail: \mbox{christian.karle@kit.edu}).}
        \thanks{B. Nuss was with the Institute of Radio Frequency Engineering and Electronics (IHE), Karlsruhe Institute of Technology (KIT), 76131 Karlsruhe, Germany. He is now with the Professorship of Microwave Sensors and Sensor Systems, Technical University of Munich, 80333 Munich, Germany. (e-mail: {benjamin.nuss@tum.de}).}
}


\maketitle

\begin{abstract} In orthogonal frequency-division multiplexing-based radar and integrated sensing and communication systems, the sensing range is traditionally limited by the round-trip time corresponding to the cyclic prefix duration. Targets whose echoes arrive after this duration induce intersymbol interference (ISI) and associated intercarrier interference (ICI), which significantly degrade detection performance, elevate the interference-noise floor in the radar image, and reduce the useful signal power due to window mismatch. Existing methods face a trade-off between recovering useful signal and suppressing interference, particularly in multi-target scenarios. This paper proposes two frameworks to resolve this dilemma, offering a flexible trade-off between computational cost and target detection performance. 
First, a signal model is derived, demonstrating that ISI and ICI-oriented interference often dominates thermal noise in high-dynamic-range scenarios. To combat the ISI and ICI-based interference-noise floor increase, joint-interference cancellation with coherent compensation is proposed. This approach is an efficient evolution of the successive-interference cancellation algorithm, utilizing high-precision chirp Z-transform estimation and frequency-domain coherent compensation to recover weak distant targets. For scenarios requiring maximum precision, the full reconstruction-based sliding window scheme is presented, which shifts the receive window to capture optimal signal energy while performing full-signal reconstruction for all detected targets. Numerical results show that both methods outperform state-of-the-art benchmarks.
\end{abstract}
\IEEEpubidadjcol
\section{Introduction}
\IEEEPARstart{I}{ntegrated} sensing and communication (ISAC) has emerged as a pivotal technology for \ac{6G} cellular networks \cite{chafii}, promising to unify sensing and data transmission into a single hardware \cite{mandelli_ISAC}. 
Among various waveform candidates, \ac{OFDM} presents itself as a robust candidate for \ac{ISAC} applications \cite{wild}, offering key advantages such as spectral efficiency and flexibility, alongside native support within existing frameworks including \ac{5G NR} and IEEE 802.11ad \cite{Zu_ISAC_challenges}.
The compatibility of OFDM with current infrastructure allows ISAC systems to deploy sensing capabilities without significant hardware changes.
Furthermore, recent \ac{3GPP} \ac{RAN} agreements have established that OFDM will be retained as the foundational downlink waveform for \ac{6G}, reaffirming its critical role in next-generation standardizations \cite{3gpp_ran109_6g_waveform}.

The dual functionality of \ac{OFDM} comes with a fundamental design conflict in waveform: the duration of the \ac{CP}. In communication systems, the CP acts as a guard interval that prevents \ac{ISI} by absorbing multipath components within its duration. It also enables circular convolution, which allows one-tap frequency-domain channel equalization. However, since the \ac{CP} does not carry any new information, it introduces overhead. To minimize this overhead and maintain a high spectral efficiency, the \ac{CP} is typically kept as short as possible, since \ac{CP} reduces the time available for payload transmission, thereby directly impacting the throughput of the system.

In contrast, radar sensing requires the detection of targets at ranges that may correspond to delays significantly exceeding the \ac{CP} duration. 
When the round-trip delay of a target echo exceeds the \ac{CP} length, two detrimental effects occur: (i) \ac{ISI} arises as the previous \ac{OFDM} symbol spills into the current detection window, and (ii) \ac{ICI} destroys the subcarrier orthogonality required for accurate parameter estimation. 
Standard \ac{OFDM} radar processing assumes that the maximum delay for echoes is strictly limited by the \ac{CP} duration, which determines the \ac{ISI}-free range. 

To overcome this limitation, recent research has diverged into two primary methodologies: \ac{CC} and \ac{SIC}. 
The \ac{CC} approach, exemplified by \cite{wang_CC}, attempts to recover the \ac{ICI}-based processing gain loss by adding the captured signal for the next symbol to the current symbol. 
While \ac{CC} effectively improves the \ac{SINR} for a single distant target, it exhibits severe drawbacks in multi-target scenarios. 
As demonstrated in \cite{geiger_MTCC}, conventional coherent compensation techniques generate strong interference and raise the interference-noise floor level in multi-target cases. 
Conversely, the \ac{SIC}-based approaches proposed in \cite{Li_SIC} focus on estimating and subtracting these interference terms iteratively to clean the radar image. 
While \ac{SIC} excels at lowering the interference-noise floor, it faces a detection threshold dilemma: it relies on the initial detection of a target to estimate interference effects for cancellation. 
If a distant target suffers from significant processing gain loss due to window mismatch, it may remain below the noise floor in the initial stage, rendering the \ac{SIC} algorithm unable to detect or cancel it.

Recently, sliding window techniques have been investigated for this scenario. 
Notably, \cite{Xu_SW} proposed a sliding window detection scheme combined with time-domain interference removal. 
While this method mitigates power degradation, it typically reconstructs signals only within the \ac{ISI}-free region for each window step. 
This partial reconstruction can lead to the late removal of interference effects, potentially preventing target detection in certain scenarios. 

Additionally, signal reconstruction techniques have been proposed in the literature to address off-grid targets and \ac{ICI} suppression.
Notably, a low-complexity super-resolution algorithm is developed in \cite{Su2025} that models the range-Doppler map via 2D local surface fitting to suppress \ac{ICI} and off-grid effects.
While this approach effectively mitigates spectral leakage, it formulates the parameter estimation as an optimization problem, requiring iterative numerical solvers (e.g., Nelder-Mead \cite{Lagarias_Nelder_Mead} or Trust-Region \cite{Richard_trust_region} methods).
Consequently, the computational cost scales with the number of detected targets and the required convergence iterations.
In contrast, our proposed framework eliminates the need for iterative fitting and any numerical solver. By leveraging the computationally efficient \ac{CZT} \cite{CZT_rabiner}, we achieve precise off-grid target detection and cancellation.

Furthermore, a critical oversight in the existing literature \cite{Li_SIC, Xu_SW} is the unrealistic assumption that targets are located on the range-Doppler grid. This would allow neglecting the necessity of windowing functions (e.g., Hamming or Chebyshev) used in practice to suppress sidelobes. 
In realistic scenarios, the absence of windowing leads to high sidelobes that can mask some of the targets, while off-grid targets introduce spectral leakage that severely degrades the accuracy of standard reconstruction-based cancellation algorithms.

This work presents a robust framework for extending the sensing range of \ac{OFDM}-based \ac{ISAC} systems under practical non-idealities, including off-grid target locations and windowing effects. Two complementary algorithms are introduced, enabling a trade-off between computational complexity and sensing capability. The main contributions are summarized as follows:

\begin{itemize}
    \item \textbf{Interference analysis under practical noise limitations:} 
    While existing models capture the effects of excessive delay \cite{wang_CC, geiger_MTCC, Li_SIC, Xu_SW}, a comprehensive analysis is formulated in this paper incorporating windowing and off-grid targets. By explicitly comparing interference power against thermal and quantization noise limits for practical systems, it is analytically shown that \ac{ISI} and \ac{ICI} from strong distant targets are the primary bottleneck in high-dynamic-range scenarios, establishing the necessity for the mitigation schemes.   
    
    \item \textbf{Joint-interference cancellation with coherent compensation algorithm:} An enhanced interference cancellation algorithm is proposed by embedding \ac{FDCC} into a joint interference cancellation scheme. This approach, named \ac{JIC-CC}, enables recovery of weak distant targets by restoring processing gain using \ac{FDCC}, while precise cancellation of off-grid strong interferers is ensured through high-resolution \ac{CZT}-based parameter estimation.
    
    \item \textbf{Full reconstruction-based sliding window algorithm:} A high-precision \ac{FR-SW} technique is developed for scenarios requiring maximum dynamic range. Unlike prior approaches such as \cite{Xu_SW} that partially mitigate interference within the \ac{ISI}-free range, \ac{FR-SW} adaptively shifts the receive window to capture optimal signal energy and performs full reconstruction of all detected targets, effectively minimizing residual interference and revealing weak targets.
    
    \item \textbf{Performance-complexity trade-off analysis:} A comparative complexity analysis is provided that evaluates both computational cost and hardware memory requirements. It is shown that while \ac{FR-SW} achieves superior interference cancellation and target detectability, \ac{JIC-CC} offers competitive performance with substantially reduced computational complexity and lower buffering demands, offering a practical option for real-time or memory-constrained systems.
\end{itemize}

The remainder of this paper is organized as follows. Section~\ref{sec:sys_model} establishes the \ac{OFDM}-\ac{ISAC} system model and analytically derives the impact of excessive delay on interference and useful signal power. Section~\ref{sec:signal_model} reviews the mathematical model with \ac{ISI} and \ac{ICI}. Section~\ref{sec:proposed_methods} introduces the proposed interference cleaning frameworks, including the \ac{JIC-CC} and the \ac{FR-SW} algorithms. Section \ref{sec:sim_res} presents the computational complexity analysis and provides comprehensive simulation results comparing the proposed methods against existing mitigation techniques and benchmarks. Section~\ref{sec:valid} verifies the proposed algorithms via verification measurements. 
Finally, Section~\ref{sec:conc} concludes the paper. 

\section{System Model and Problem Formulation}
\label{sec:sys_model}
This study considers a monostatic \ac{CP}-\ac{OFDM}-\ac{ISAC} system transmitting an \ac{OFDM} frame composed of $M$ \ac{OFDM} symbols, each consisting of $N$ subcarriers. The baseband time-domain transmit signal of the $m\text{th}$ \ac{OFDM} symbol is
\begin{equation}
x_m(t) = \sqrt{\frac{P_\text{tx}}{N}} \sum_{k=0}^{N-1} X_m(k) \, \mathrm{e}^{\mathrm{j} 2\pi k \Delta f t} u(t),
\label{eq:xm}
\end{equation}
where $X_m(k) \in \mathbb{C}$ denotes the unit power complex-valued data symbol modulated onto the $k$th subcarrier of the $m$th OFDM symbol and entry of $\mathbf{X}\in \mathbb{C}^{N \times M}$, $P_\text{tx}$ is the total average transmit power per OFDM symbol, $\Delta f$ denotes the subcarrier spacing\footnote{For notational convenience, the baseband signal in this mathematical model is defined over the frequency interval $[0, B]$, effectively centered at $B/2$. In actual hardware realizations, the complex discrete-time baseband signal is generated symmetrically around $0$\,Hz before upconversion. Consequently, applying this practical symmetric representation directly to the radar signal processing framework detailed in Section~\ref{sec:signal_model} would necessitate appropriate subcarrier reindexing and slight modifications to the subsequent mathematical formulations. Furthermore, the explicit transformation of these time-discrete symbols into a continuous-time waveform via a \ac{DAC} is omitted for brevity.}.
The function $u(t)$ represents a rectangular pulse window that incorporates the \ac{CP}
\begin{equation}
u(t) =
\begin{cases}
1, & -T_{\mathrm{cp}} \leq t < T_\text{d} \\
0, & \text{otherwise},
\end{cases}
\label{eq:ut}
\end{equation}
where $T_{\text{cp}}$ denotes the cyclic prefix duration, while
$T_{\text{d}}$ represents the data-carrying part of the \ac{OFDM} symbol excluding the
\ac{CP}. Accordingly, the total duration of one \ac{OFDM} symbol is
$T = T_{\text{cp}} + T_{\text{d}}$.
The transmit signal $x(t)$ corresponds to a frame consisting of $M$ \ac{OFDM} symbols and is then given by
\begin{align}
x(t) &= \sum_{m=0}^{M-1} x_m(t-mT) \nonumber \\
     &= \sqrt{\frac{P_\text{tx}}{N}} \sum_{m=0}^{M-1} \sum_{k=0}^{N - 1} X_m(k) \, 
     \mathrm{e}^{\mathrm{j} 2\pi k \Delta f (t - mT)} u(t - mT).
\label{eq:x_total}
\end{align}

\begin{figure}
    \centering
    \includegraphics[width=1\linewidth]{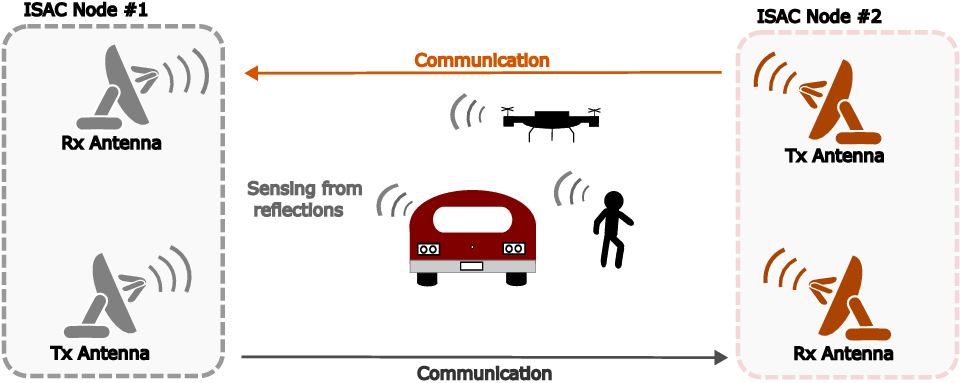}
    \caption{Considered ISAC system model.}
    \label{fig:system_model}
    \vspace{-2mm}
\end{figure}

The sensing channel is modeled with $H$ targets present in the environment, alongside multiple \ac{ISAC} nodes, as illustrated in Fig.~\ref{fig:system_model}. It is assumed that the communication signal from \ac{ISAC} Node \#2 is perfectly cancelled prior to sensing processing at the receiver of \ac{ISAC} Node \#1. 
This assumption isolates the sensing reflections and enables a focused analysis of radar, i.e., sensing performance.

Under this assumption, the received sensing signal can be expressed as
\begin{equation}
y(t) = \sum_{h=0}^{H-1} \alpha_h x(t-\tau_h) \mathrm{e}^{\mathrm{j} 2\pi f_{\text{D},h} t} + w(t),
\label{eq:yt}
\end{equation}
where $\alpha_h \in \mathbb{C}$ is complex attenuation factor of target $h$, $\tau_h$ is the delay of target $h$, $f_{\text{D},h}$ denotes the Doppler frequency for target $h$, and $w(t)$ denotes \ac{AWGN} \mbox{$w(t) \sim \mathcal{CN}(0, \sigma^2)$} with noise power $\sigma^2$. For completeness, $|\alpha_h|$ follows from the \cite{canan_aydogdu}, \cite{lucas_overview_schemes},
\begin{equation}
|a_h| = \sqrt{ \frac{ G_{\text{tx}} \, G_{\text{rx}} \sigma_{\text{RCS},h} \, \lambda^2 }{ (4\pi)^3 \, R_h^4} },
\label{eq:ap}
\end{equation}
where $G_\text{tx}$ and $G_\text{rx}$ are the transmitter and receiver antenna gains, respectively, $\sigma_{\text{RCS},h}$ is the radar cross section of target $h$, $R_h$ represents the distance to target $h$, and $\lambda = c/f_\text{c}$ is the wavelength of the transmitted signal, with $c$ denoting the speed of light.
 
In an ideal scenario, it is assumed that (i) targets are perfectly separated by the resolution limits, (ii) target delays are bounded by the \ac{CP} duration avoiding \ac{ISI} and \ac{ICI}, and (iii) the \ac{ADC} resolution is high enough, meaning quantization noise is negligible and thermal noise is the sole limiting factor. Under these ideal assumptions, the resulting radar image exhibits the $h$th target as a distinct signal peak. The corresponding \ac{SNR} for target $h$, denoted as $\text{SNR}_{\text{image},h}^{\text{ideal}}$, can be computed with respect to the post-processing noise variance as \cite{lucas_overview_schemes}
\begin{equation}
    \text{SNR}^\text{ideal}_{\text{image},h} = \frac{P_{\text{tx}} \, G_{\text{tx}} \, G_{\text{rx}} \, \sigma_{\text{RCS},h} \, \lambda^2 \, G_\text{p}}{(4\pi)^3 \, R_h^4 \, k_\text{B} \, B \, T_{\text{th}} \, \text{NF}},
\label{eq:SNR}
\end{equation}
with processing gain $G_\text{p}$. The term $k_\text{B}$ represents Boltzmann's constant, $T_{\text{th}}$ is the standard room temperature in Kelvin, and $\text{NF}$ is the overall receiver noise figure. 
Consequently, the product $k_\text{B} \, B \, T_{\text{th}} \, \text{NF}$ in the denominator accounts for the effective power of the sampled version of the \ac{AWGN} $w(t)$.
However, this idealized formula fails to capture the performance bottlenecks of practical systems. In reality, the dynamic range is severely constrained by quantization noise, as well as \ac{ISI} and \ac{ICI} from targets exceeding the \ac{CP} duration. To quantify these effects, the expression in \eqref{eq:SNR} must be extended.

For \ac{OFDM}-based radar systems, the \ac{CP} duration determines the maximum \ac{ISI}-free range
\begin{equation}
    R_\text{max,ISI} = \frac{c \cdot T_\text{cp}}{2}.
\end{equation}
The targets that are in this \ac{ISI}-free region do not induce any \ac{ISI} and \ac{ICI} due to delay. 
On the other hand, target motion induces a Doppler shift that inherently results in \ac{ICI}. To ensure that subcarrier orthogonality is approximately preserved and the resulting \ac{ICI} remains negligible, the Doppler shift is typically constrained to $f_\text{D} < 0.1\,\Delta f$, which defines the maximum \ac{ICI}-limited velocity $v_{\text{max,ICI}}$ \cite{lucas_overview_schemes, Nuss2018}.

\begin{figure}[t]
	\centering
	\includegraphics[width=0.8\columnwidth]{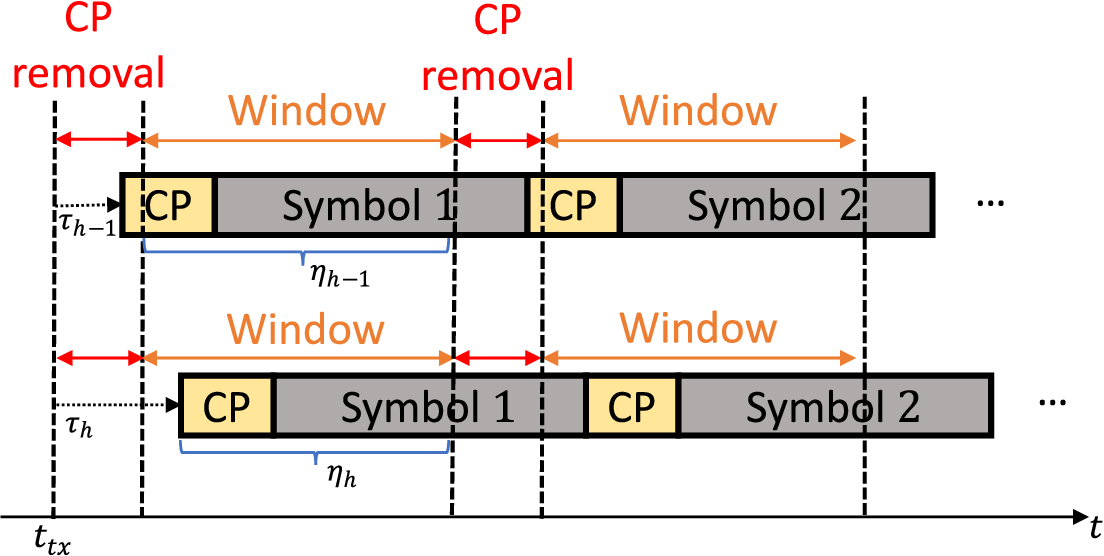}
	\caption{Received OFDM signal for two different targets, upper is for the target in the ISI free region and lower is for the non-ISI free region.}
	\label{fig:CP_figure}
\end{figure}
While this velocity limit is necessary to mitigate drastic Doppler-induced interference, the primary structural constraint regarding \ac{ISI} is imposed by the \ac{CP} duration. This is visualized in Fig.~\ref{fig:CP_figure}, which illustrates the reception timing relative to the transmit time $t_\text{tx}$ for two distinct scenarios. For target $h-1$, the round-trip delay is shorter than the \ac{CP} duration, therefore, the received echo remains fully aligned within the \ac{ISI}-free region, avoiding delay-induced interference. In contrast, the echo from target $h$ arrives with a delay exceeding the \ac{CP} duration. This misalignment destroys subcarrier orthogonality, resulting in both \ac{ISI} and \ac{ICI}. Consequently, the target $h$ suffers from a reduction in effective useful signal power due to window mismatch.
This scenario leads to both \ac{ISI} and \ac{ICI}, resulting in interference and loss in useful signal power.
To quantify this power loss, the captured signal fraction for target $h$ is defined as
\begin{equation}
\eta_h = \text{max}\left (0,\text{min}\left(1 - \frac{\tau_h - T_{\text{cp}}}{T_\text{d}}, 1\right)\right).
\label{eq:eta_h}
\end{equation}
Accounting for the impact of window mismatch and interference, the actual post-processing \ac{SINR} for target $h$ is given by
\begin{equation}
\text{SINR}^\text{actual}_{\text{image},h} = \frac{P_{\text{tx}}  G_{\text{tx}}  G_{\text{rx}}  \sigma_{\text{RCS},h}  \lambda^2  G_\text{p}  \eta^2_h}{(4\pi)^3  R_h^4  P_{n}^\text{dom} },
\label{eq:SNR_act}
\end{equation}
where $P_n^\text{dom}$ denotes the dominant noise or interference term over thermal noise, quantization noise, and interference limiting the system dynamic range.

In practical monostatic \ac{ISAC} systems, a strong self-interference arises due to the spillover between the transmit and receive chains. 
Since this coupled signal is typically much stronger than the reflected target echoes \cite{lucas_overview_schemes}, it dictates the required full-scale range of the \ac{ADC} to prevent signal clipping. Consequently, the quantization noise floor is characterized by the \ac{SQNR}, which must account for the \ac{PAPR} of the \ac{OFDM} waveform. 
Specifically, the \ac{SQNR} can be expressed as \cite{widrow_quantization}
\begin{equation}
\text{SQNR} = 6.02N_{\text{bit}} + 10\log_{10}(3F), 
\label{eq:dr}
\end{equation}
where $N_{\text{bit}}$ denotes the number of bits used in the \ac{ADC} and $F$ represents the ratio of the average signal power to the squared peak amplitude. In this study, the quantization noise power is determined relative to the peak spillover level, ensuring it is accurately modeled alongside thermal noise and interference.
 
Beyond hardware-induced thermal noise and quantization noise limits, the system is also limited by excessive delay-induced interference. Specifically, the interference power resulting from a target that is outside the \ac{ISI}-free region can be calculated as
\begin{equation}
    \text{P}_{\text{int},h} = \frac{P_{\text{tx}} \, G_{\text{tx}} \, G_{\text{rx}} \, \sigma_{\text{RCS},h} \, \lambda^2 \, (1-\eta^2_h)}{(4\pi)^3 \, R_h^4},
\label{eq:int_power_p}
\end{equation}
where $\text{P}_{\text{int},h}$ denotes the interference power level resulting from target $h$.
\begin{table}
\vspace{-0.2cm}
    \renewcommand{\arraystretch}{1.5}
    \setlength{\arrayrulewidth}{.1mm}
    \setlength{\tabcolsep}{4pt}   		
    \centering   \captionsetup{width=43pc,justification=centering,labelsep=newline}
    \caption{\textsc{Considered Simulation Parameters}}
    \label{tab:sim_pars}
    \begin{tabular}{lll}
        \toprule
        \textbf{Parameters} & \textbf{Symbol} & \textbf{Value} \\
        \midrule
        Carrier frequency  & $f_\mathrm{c}$ & \qty{3.5}{\giga\hertz} \\
        Frequency bandwidth & $B$ & \qty{200}{\mega\hertz} \\
        Number of subcarriers & $N$ & \num{6652} \\
        Cyclic prefix length & $N_{\mathrm{cp}}$ & $458$ \\
        OFDM symbols per frame & $M$ & \num{280}  \\
        Transmit power & $P_\mathrm{tx}$ & \qty{49}{dBm}  \\
        Antenna gain & $G_\mathrm{tx}$, $G_\mathrm{rx}$ & \qty{25.8}{dBi} \\
        Noise Figure & $\mathrm{NF}$ & \qty{8}{dB} \\
        \bottomrule
    \end{tabular}
    \vspace{-0.2cm}
\end{table} 
 \begin{figure}[t]
    \centering
    % --- PSFRAG REPLACEMENTS (Optional: Add axis number replacements here if needed) ---
    \psfrag{Range (m)}[c][c]{\scriptsize Range (m)}
    \psfrag{Power (dBm)}[c][c]{\scriptsize Power (dBm)}
    \psfrag{0}[c][c]{\scriptsize $0$}
    \psfrag{1000}[c][c]{\scriptsize $1000$}
    \psfrag{2000}[c][c]{\scriptsize $2000$}
    \psfrag{3000}[c][c]{\scriptsize $3000$}
    \psfrag{4000}[c][c]{\scriptsize $4000$}
    \psfrag{5000}[c][c]{\scriptsize $5000$}
    \psfrag{100}[c][c]{\scriptsize $100$}
    \psfrag{80}[c][c]{\scriptsize $80$}
    \psfrag{60}[c][c]{\scriptsize $60$}
    \psfrag{40}[c][c]{\scriptsize $40$}
    \psfrag{20}[c][c]{\scriptsize $20$}
    \psfrag{-100}[c][c]{\scriptsize -$100$}
    \psfrag{-80}[c][c]{\scriptsize -$80$}
    \psfrag{-60}[c][c]{\scriptsize -$60$}
    \psfrag{-40}[c][c]{\scriptsize -$40$}
    \psfrag{-20}[c][c]{\scriptsize -$20$}
    % --- INCLUDE GRAPHICS ---
    \includegraphics[width=0.9\columnwidth]{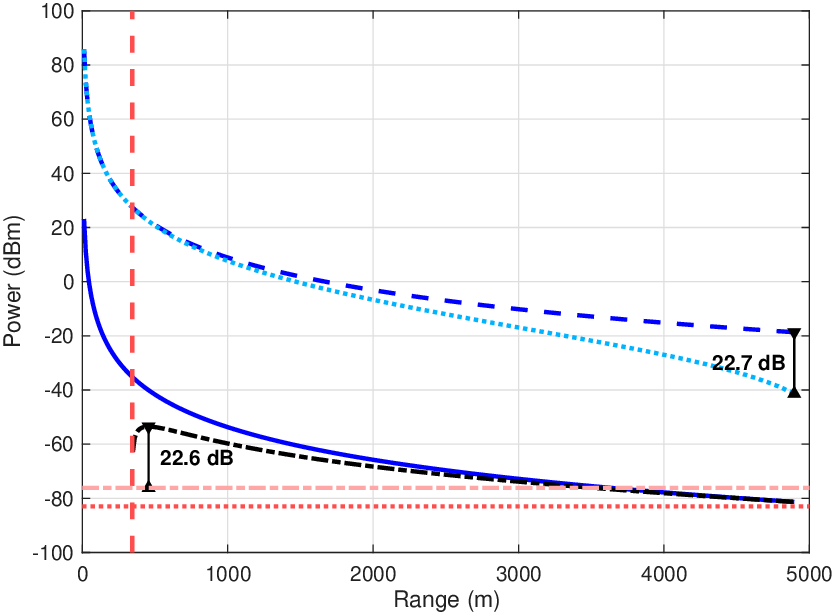}    
    % --- CAPTION WITH MANUAL LEGEND ---
    \caption{Received power profile versus range for a target with 20 dBsm \ac{RCS}. 
    The effective received power is shown as 
    ({\color[rgb]{0,0,1}\textbf{\rule[0.5ex]{1.5em}{1.25pt}}}). 
    The maximum \ac{ISI} range $R_{\text{max,ISI}}$ is indicated by 
    ({\color[rgb]{1,0,0}\textbf{\rule[0.5ex]{0.75em}{1.25pt}\hspace{0.2em}\rule[0.5ex]{0.75em}{1.25pt}}}). 
    Noise floors are depicted as follows: 
Thermal noise is shown as
({\color[rgb]{1,0,0}\textbf{\rule[0.5ex]{0.2em}{1.25pt}\hspace{0.15em}\rule[0.5ex]{0.2em}{1.25pt}\hspace{0.15em}\rule[0.5ex]{0.2em}{1.25pt}}}),
and 12-bit quantization noise as
({\color[rgb]{1,0.5,0.5}\textbf{\rule[0.5ex]{0.75em}{1.25pt}\hspace{0.15em}\rule[0.5ex]{0.2em}{1.25pt}}}).
The interference power is indicated by
({\color[rgb]{0,0,0}\textbf{\rule[0.5ex]{0.75em}{1.25pt}\hspace{0.15em}\rule[0.5ex]{0.2em}{1.25pt}}}).
The effective received signal, including processing gain, is represented by
({\color[rgb]{0,0,1}\textbf{\rule[0.5ex]{0.75em}{1.25pt}\hspace{0.2em}\rule[0.5ex]{0.75em}{1.25pt}}}),
and the peak power in the radar image by
({\color[rgb]{0,0.7,1}\textbf{\rule[0.5ex]{0.2em}{1.25pt}\hspace{0.15em}\rule[0.5ex]{0.2em}{1.25pt}\hspace{0.15em}\rule[0.5ex]{0.2em}{1.25pt}}}).}
    \label{fig:motivation}
    \vspace{-0.2cm}
\end{figure}

To formulate the problem and motivate the necessity of the proposed interference cancellation frameworks prior to the detailed signal model derivations, a preliminary numerical system-level analysis is conducted. To identify the dominant noise or interference contribution in practical systems among thermal, quantization, and interference in practical systems, the system parameters reported in \cite{mandelli_ISAC} and \cite{geiger_MTCC} are considered, which are summarized in Table~\ref{tab:sim_pars}.
For an aggregate Tx-Rx isolation of \num{60}~dB, which is achievable through a combination of physical antenna separation and analog self-interference cancellation \cite{dinesh2013}, the quantization noise level is computed based on the spillover level and \eqref{eq:dr}.

Using the parameters in Table~\ref{tab:sim_pars}, an example scenario is considered in which a target with \qty{20}{dBsm} \ac{RCS} is generated, and its range is swept up to the maximum unambiguous range. The corresponding results are shown in Fig.~\ref{fig:motivation}.
When a strong target (e.g., a automobile or truck with \qty{20}{dBsm} \ac{RCS} \cite{Richards2005}) moves outside the \ac{ISI}-free range, it introduces significant interference as illustrated in Fig.~\ref{fig:motivation}. This interference results in an interference-induced floor in the radar image that exceeds both the quantization noise and the thermal noise levels. As annotated in Fig.~\ref{fig:motivation}, the noise level difference between the interference noise and the larger of the thermal and quantization noise components can reach up to \qty{22.6}{dB}, depending on the range of the interferer.
Furthermore, as the target range approaches the maximum unambiguous range, the received power in the radar image decreases due to the reduced overlap between the delayed echo and the processing window of the receiver. For the considered system parameters in Table~\ref{tab:sim_pars}, this power loss can reach up to \qty{22.7}{dB}.
\begin{figure}
    \centering
    \psfrag{Interference Source RCS (dBsm)}[c][c]{\scriptsize Interferer RCS (dBsm)}
    \psfrag{Power level (dBm)}[c][c]{\scriptsize Power (dBm)}
    \psfrag{Maximum Range (km)}[c][c]{\scriptsize Maximum range (km)}
    \psfrag{(a)}[c][c]{\scriptsize (a)}
    \psfrag{(b)}[c][c]{\scriptsize (b)}
    \psfrag{0}[c][c]{\scriptsize $0$}
    \psfrag{10}[c][c]{\scriptsize $10$}
    \psfrag{20}[c][c]{\scriptsize $20$}
    \psfrag{30}[c][c]{\scriptsize $30$}
    \psfrag{-85}[c][c]{\scriptsize -$85$}
    \psfrag{-80}[c][c]{\scriptsize -$80$}
    \psfrag{-75}[c][c]{\scriptsize -$75$}
    \psfrag{-70}[c][c]{\scriptsize -$70$}
    \psfrag{-65}[c][c]{\scriptsize -$65$}
    \psfrag{-60}[c][c]{\scriptsize -$60$}
    \psfrag{-55}[c][c]{\scriptsize -$55$}
    \psfrag{-50}[c][c]{\scriptsize -$50$}
    \psfrag{-45}[c][c]{\scriptsize-$45$}
     \psfrag{-40}[c][c]{\scriptsize-$40$}
     \psfrag{0}[c][c]{\scriptsize $0$}
    \psfrag{0.5}[c][c]{\scriptsize $0.5$}
    \psfrag{1}[c][c]{\scriptsize $1$}
    \psfrag{1.5}[c][c]{\scriptsize $1.5$}
    \psfrag{2}[c][c]{\scriptsize $2$}
    \psfrag{2.5}[c][c]{\scriptsize $2.5$}
    \psfrag{3}[c][c]{\scriptsize $3$}
    \psfrag{3.5}[c][c]{\scriptsize $3.5$}
    \psfrag{4}[c][c]{\scriptsize $4$}
    \psfrag{4.5}[c][c]{\scriptsize $4.5$}
    \includegraphics[width=0.5\textwidth]{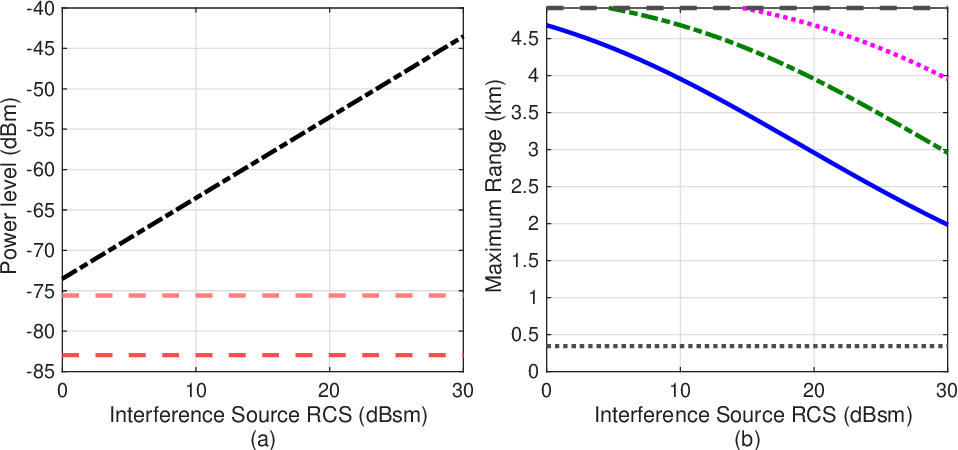}
\caption{Analysis of noise, interference levels, and maximum detectable range versus interference source RCS.  
(a) Comparison of noise and interference power levels. Interference power is represented by  
({\color[rgb]{0,0,0}\textbf{\rule[0.5ex]{0.75em}{1.25pt}\hspace{0.15em}\rule[0.5ex]{0.2em}{1.25pt}}}),  
thermal noise by  
({\color[rgb]{1,0,0}\textbf{\rule[0.5ex]{0.75em}{1.25pt}\hspace{0.2em}\rule[0.5ex]{0.75em}{1.25pt}}}),  
and quantization noise by  
({\color[rgb]{1,0.5,0.5}\textbf{\rule[0.5ex]{0.75em}{1.25pt}\hspace{0.2em}\rule[0.5ex]{0.75em}{1.25pt}}}).  
(b) Maximum detectable range for target RCS values of $0~\mathrm{dBsm}$  
({\color[rgb]{0,0,1}\textbf{\rule[0.5ex]{1.5em}{1.25pt}}}),  
$10~\mathrm{dBsm}$  
({\color[rgb]{0,0.5,0}\textbf{\rule[0.5ex]{0.75em}{1.25pt}\hspace{0.15em}\rule[0.5ex]{0.2em}{1.25pt}}}),  
and $20~\mathrm{dBsm}$  
({\color[rgb]{1,0,1}\textbf{\rule[0.5ex]{0.2em}{1.25pt}\hspace{0.15em}\rule[0.5ex]{0.2em}{1.25pt}\hspace{0.15em}\rule[0.5ex]{0.2em}{1.25pt}}}).  
The maximum unambiguous range limit is indicated by  
({\color[rgb]{0,0,0}\textbf{\rule[0.5ex]{0.75em}{1.25pt}\hspace{0.2em}\rule[0.5ex]{0.75em}{1.25pt}}}),  
and the ISI-free range limit by  
({\color[rgb]{0,0,0}\textbf{\rule[0.5ex]{0.2em}{1.25pt}\hspace{0.15em}\rule[0.5ex]{0.2em}{1.25pt}\hspace{0.15em}\rule[0.5ex]{0.2em}{1.25pt}}}).}
    \label{fig:noise_range_analysis}
    \vspace{-0.2cm}
\end{figure}

To evaluate the system limits, a worst-case scenario is analyzed where the interfering target is positioned at a range that maximizes interference power. In this configuration, the \ac{RCS} of the interferer is swept to compare the resulting noise levels, as illustrated in Fig.~\ref{fig:noise_range_analysis}(a). The results demonstrate that interference-induced noise consistently exceeds other noise sources across all considered cases. Consequently, interference is identified as the dominant factor that determines the floor level in the radar image and is utilized in \eqref{eq:SNR_act} for performance estimation. Based on the minimum target \ac{RCS} and the required \ac{SNR} threshold of \qty{17}{dB} for reliable detection of a target \cite{mandelli_ISAC}, the maximum detectable range is calculated and plotted as a function of the \ac{RCS} of interferer in Fig.~\ref{fig:noise_range_analysis}(b). It is evident that the detectable range is drastically reduced when a target or multiple targets exist beyond the \ac{ISI}-free range. This substantial degradation highlights the critical necessity for robust interference cancellation and coherent compensation mechanisms.
\section{Signal Model} 
\label{sec:signal_model}

The received signal associated with target $h$, denoted as $y^h(t)$, is analyzed under two distinct delay scenarios, as illustrated in Fig.~\ref{fig:CP_figure}. The first case considers a target whose round-trip delay $\tau_h$ is shorter than the \ac{CP} duration. As a result, the corresponding echo does not experience \ac{ISI} or \ac{ICI} due to excessive delay, as depicted in the upper part of Fig.~\ref{fig:CP_figure}. For this scenario, if $i \in \{0,\ldots, N-1\}$  is used to denote the sample index at the $m$th \ac{OFDM} symbol interval, the received samples can be written as
\begin{align}
	y_{m}^h(i) &= \sqrt{\frac{1}{N}} \sum_{k=0}^{N-1} \tilde{\alpha}_{h} \cdot X_m(k) \cdot 
	\text{e}^{\text{j} 2\pi k \Delta f (iT_s - \tau_h)} \notag \\
	&\quad \cdot \mathrm{e}^{\text{j} 2\pi f_{\text{D},h} i T_s} \cdot u(iT_s - \tau_h),
	\label{eq:y(i)_noISI}
\end{align} 
where $|\tilde{\alpha}_h|^2 = P_{\text{tx}}|\alpha_h|^2$.
Applying a \ac{DFT} on \eqref{eq:y(i)_noISI}, the resulting frequency domain signal is given by 

\begin{equation}
	Y_{m}^h (p) = \tilde{\alpha}_h X_m(p) \mathrm{e}^{\frac{-\mathrm{j} 2 \pi p \tau_h}{N T_s}} \mathrm{e}^{\mathrm{j} 2 \pi f_{\text{D},h} m T},
	\label{eq:Y(p)_noISI}
\end{equation} 
where $p \in \{0,\ldots N-1\}$.

Next, the excessive delay scenario is considered, in which the round-trip delay of target~$h$ exceeds the \ac{CP} duration. In such cases, the received samples corresponding to target~$h$ are influenced by two consecutive \ac{OFDM} symbols, resulting in both \ac{ISI} and \ac{ICI}. The received signal under this condition is expressed as~\eqref{eq:y(i)} \cite{wang_CC}, in which $w_{m}(i)$ denotes the \ac{AWGN} at the $i$th sample of the $m$th received OFDM symbol. 
\begin{figure*}[t]	
	\begin{equation}
		\begin{split}
			y_{m}^h(i) 
			&= \sqrt{\frac{1}{N}} \sum_{k=0}^{N-1} \tilde{\alpha}_h \cdot X_{m}(k) \cdot 
			\mathrm{e}^{\mathrm{j} 2\pi k \Delta f (iT_s - \tau_h)} \cdot 
			\mathrm{e}^{\mathrm{j}2\pi f_{\text{D},h} (mT + iT_s)} \cdot u(iT_s - \tau_h) \\
			&\quad + \sqrt{\frac{1}{N}} \sum_{k=0}^{N-1} \tilde{\alpha}_h \cdot X_{m{-}1}(k) \cdot 
			\mathrm{e}^{\mathrm{j} 2\pi k \Delta f (T + iT_s - \tau_h)} \cdot 
			\mathrm{e}^{\mathrm{j}2\pi f_{\text{D},h} (mT + iT_s)} \cdot u(T + iT_s - \tau_h) + w_{m}(i)
		\end{split}
		\label{eq:y(i)}
	\end{equation}
	\vspace{-0.65cm}
    \rule{\textwidth}{0.4pt}
\end{figure*}
\begin{figure*}[t]
	\begin{equation}
		\begin{split}
			Y^h_{m} (p) &= \left(1 - \frac{N_h - N_\text{cp}}{N}\right)\tilde{\alpha}_h X_{m}(p) \mathrm{e}^{\frac{-\mathrm{j} 2 \pi p \tau_h}{N T_s}} \mathrm{e}^{\mathrm{j} 2 \pi f_{\text{D},h} m T} + \frac{1}{N} \sum_{k=0}^{N - 1} \left[ \tilde{\alpha}_h X_{m-1}(k) \mathrm{e}^{\frac{\mathrm{j} 2 \pi k (T_\text{cp} - \tau_h)}{N T_s}} \left( \sum_{i=0}^{N_h - N_\text{cp} - 1} \mathrm{e}^{\frac{j 2 \pi (k-p) i}{N}} \right)\right]\\
			&\quad \mathrm{e}^{\mathrm{j} 2 \pi f_{\text{D},h} m T} 
			 + \frac{1}{N} \sum_{k=0,k\neq p}^{N - 1} \left[ \tilde{\alpha}_h X_{m}(k) \mathrm{e}^{\frac{-\mathrm{j} 2 \pi k \tau}{N T_s}} \left( \sum_{i=N_h - N_\text{cp}}^{N - 1} \mathrm{e}^{\frac{\mathrm{j} 2 \pi (k-p) i}{N}} \right)\right] \mathrm{e}^{\mathrm{j} 2 \pi f_{\text{D},h} m T} + W_{m} (p).
		\end{split}
	\label{eq:Y(p)}
    \end{equation}
    \rule{\textwidth}{0.4pt}
\end{figure*}

Following the analysis in~\cite{wang_CC}, the average power of the useful signal can be calculated under the assumption of low Doppler shifts, i.e., $f_{\text{D},h} < \frac{\Delta f}{10}$. By applying the \ac{DFT} to~\eqref{eq:y(i)}, the resulting expression can be computed as in~\eqref{eq:Y(p)}. 
By rearranging \eqref{eq:Y(p)}, the received signal can be expressed as \cite{Li_SIC}
	\begin{equation}
		\begin{split}
			{Y}_{m}^h(p) &= \underbrace{\tilde{\alpha}_h X_m(p) \text{e}^{-\text{j} 2\pi p \Delta f \tau_h} \text{e}^{\text{j} 2\pi m f_{\text{D},h} T_s}}_{\text{Y}_{m,h}^\text{free}} \\
			& + \underbrace{\tilde{\alpha}_h \mathrm{e}^{\mathrm{j} 2\pi (m{-}1) f_{\text{D},h} T_s} \sum_{k=0}^{N-1} X_{m{-}1}(k) \mathrm{e}^{\mathrm{j} 2\pi k \Delta f (T_{cp} {-} \tau_h)} \phi_{p,k}^h}_{\text{Y}_{m,h}^\text{ISI}} \\
			& - \underbrace{\tilde{\alpha}_h \mathrm{e}^{\mathrm{j} 2\pi m f_{\text{D},h} T_s} \sum_{k=0}^{N-1} X_m(k) \mathrm{e}^{-\mathrm{j} 2\pi k \Delta f \tau_h} \phi_{p,k}^h}_{\text{Y}_{m,h}^\text{ICI}},
		\end{split}
		\label{eq:expr_Li}
	\end{equation}
where $\phi_{p,k}^h = \frac{1}{N}\sum_{i=N_h - N_\text{cp}}^{N - 1} \mathrm{e}^{\frac{\mathrm{j} 2 \pi (k-p) i}{N}}$ represents the interference component, which also denotes the element at row $p$ and column $k$ of the interference matrix $\mathbf{\Phi}_h \in \mathbb{C}^{N \times N}$.

Let the steering vectors for the delay, Doppler, and Jacobian matrix be defined as \cite{Li_SIC}
\begin{align}
    \label{eq:steering_vec_range}
\mathbf{b}(\tau_h) &= \left[1, \mathrm{e}^{-\mathrm{j}\pi \Delta f \tau_h}, \dots, \mathrm{e}^{-\mathrm{j}2\pi (N-1)\Delta f \tau_h} \right]^\text{T}, \\
    \label{eq:steering_vec_doppler}
    \mathbf{c}(f_{\text{D},h}) &= \left[1, \mathrm{e}^{\mathrm{j}2\pi f_{\text{D},h} T}, \dots, \mathrm{e}^{\mathrm{j}2\pi (M-1)f_{\text{D},h} T} \right]^\text{T}, \\
    \label{eq:shift_matrix}
    \mathbf{J} &= \begin{bmatrix} 
        \mathbf{0}_{(M-1)\times 1} & \mathbf{I}_{M-1} \\
        0 & \mathbf{0}_{1 \times (M-1)}
    \end{bmatrix},
\end{align}
where $\mathbf{I}_{M-1}$ is the identity matrix, $\mathbf{0}_{(M-1)\times 1}$ and $\mathbf{0}_{1 \times (M-1)}$ are zero column and row vectors with size $M-1$, respectively. Based on these steering vectors, and the Jacobian matrix, and~\eqref{eq:expr_Li}, the signal contribution from each target in the frequency domain can be expressed as
\begin{equation}
    \begin{split}
        \mathbf{Y}_h (\tilde{\alpha}_h,\tau_h, f_{\text{D},h}) &= \tilde{\alpha}_h \Bigg[ \left( \mathbf{b}(\tau_h)\mathbf{c}^\text{T}(f_{\text{D},h}) \odot \mathbf{X} \right) \\ 
        & + \mathbf{\Phi}_{h} \left( \mathbf{b}(\tau_h - T_{\mathrm{cp}})\mathbf{c}^\text{T}(f_{\text{D},h}) \odot \mathbf{X} \right) \mathbf{J} \\
        & + \mathbf{\Phi}_{h} \left( \mathbf{b}(\tau_h)\mathbf{c}^\text{T}(f_{\text{D},h}) \odot \mathbf{X} \right) \Bigg].
    \end{split}
    \label{eq:Y_h_in_matrix_form}
\end{equation}
where $\odot$ denotes the Hadamard product. Finally, the received frequency-domain frame without noise $\mathbf{Y} \in \mathbb{C}^{N \times M}$ can be expressed as the superposition of all target reflections as
\begin{equation}
    \mathbf{Y} = \sum\limits_{h=0}^{H-1}\mathbf{Y}_h (\tilde{\alpha}_h,\tau_h,f_{\text{D},h}).
\end{equation}
The estimated channel matrix $\mathbf{H}$ with zero-forcing equalization and 2D windowing is given by

\begin{equation}
    \mathbf{H} = \left( \left[ \mathbf{Y} \oslash \mathbf{X} \right] \right) \odot \mathbf{H}_{\text{window}},
    \label{eq:channel_est}
\end{equation}
where $\oslash$ denotes the element-wise division. 
The 2D windowing matrix is defined as the outer product $\mathbf{H}_{\text{window}} = \mathbf{w}_r \mathbf{w}_d^T$, where $\mathbf{w}_r \in \mathbb{R}^N$ and $\mathbf{w}_d \in \mathbb{R}^M$ are the windowing coefficient vectors for the delay and Doppler dimensions.

Finally, the range-velocity image $\mathbf{I}_\text{rv}$ is obtained via standard range-Doppler processing

\begin{equation}
\mathbf{I}_\text{rv} = \mathcal{F}_M \left\{ \mathcal{F}_N^{-1} \{ \mathbf{H} \} \right\} \mathbf{\Pi}_M,
    \label{eq:rd_processing}
\end{equation}

\noindent where $\mathcal{F}_N^{-1}$, $\mathcal{F}_M$, and $\mathbf{I}_\text{rv} \in \mathbb{C}^{N \times M}$  denote the energy preserving \ac{IDFT} along the subcarriers, the \ac{DFT} along the symbols, and the complex radar image, respectively.  The matrix $\mathbf{\Pi}_M \in \{0,1\}^{M \times M}$ is a block permutation matrix that
ensures the zero-velocity component is aligned to the center of the image. Assuming $M$ is even, $\mathbf{\Pi}$ is defined as

\begin{equation}
    \mathbf{\Pi}_M = 
    \begin{bmatrix}
        \mathbf{0}_{M/2} & \mathbf{I}_{M/2} \\
        \mathbf{I}_{M/2} & \mathbf{0}_{M/2}
    \end{bmatrix},
    \label{eq:permutation_matrix}
\end{equation}
where $\mathbf{0}_{M/2}$ denote the zero matrix with size $M/2$.

Following a coarse target detection in the range-velocity image $|\mathbf{I}_\text{rv}|^2$ by the \ac{CFAR} algorithm, a 2D \ac{CZT} can be employed to refine the range and velocity estimates. Unlike the standard \ac{FFT}, which is limited to a fixed grid resolution of $\Delta f = f_\text{s}/N$ (where $f_\text{s}$ denotes the sampling frequency), the \ac{CZT} allows for the evaluation of the Z-transform along a spiral contour. This effectively zooms into a specific spectral \ac{ROI} with an arbitrary resolution factor $L$.

Let $(\hat{n}_h, \hat{m}_h)$ denote the coarse range and Doppler indices of a detected peak for target $h$ where $\hat{n}_h \in \{0, \dots, N-1\}$ and $\hat{m}_h \in \{0, \dots, M-1\}$. A local search window of width $B_{\text{roi}}$ bins is defined around these indices. The refinement procedure is carried out by range interpolation followed by Doppler interpolation.
In the first stage, \ac{CZT} is applied along the first dimension of the pre-processed, windowed channel response $\mathbf{H}$. The zoom operation is controlled by the starting contour point $A_\text{r}$ and the contour step scalar $W_\text{r}$, which are defined as

\begin{equation}
    A_\text{r} = \mathrm{e}^{-\mathrm{j} \frac{2\pi}{N} \left(\hat{n}_h - \frac{B_{\text{roi}}}{2}\right)}, \quad 
    W_\text{r} = \mathrm{e}^{\mathrm{j} \frac{2\pi}{L\cdot N}}.
    \label{eq:czt_range_params}
\end{equation}
The intermediate range-refined matrix $\mathbf{Z}_\text{r}$ is computed by evaluating the \ac{CZT} along the defined contour

\begin{equation}
    \mathbf{Z}_\text{r}(k, m) = \sum_{n=0}^{N-1} \mathbf{H}(n, m) A_\text{r}^{n} W_\text{r}^{nk},
    \label{eq:czt_range_sum}
\end{equation}

\noindent  where $N_\text{czt} = B_\text{roi}\cdot L$ and $k \in \{0, \dots, N_\text{czt}-1\}$.
Next, the \ac{CZT} is applied along the second dimension of $\mathbf{Z}_\text{r}$ to refine the velocity estimate. The Doppler contour parameters are defined similarly around the coarse Doppler index $\hat{m}_h$ as

\begin{equation}
    A_\text{d} = \mathrm{e}^{\mathrm{j} \frac{2\pi}{M} \left(\hat{m}_h - \frac{M+B_{\text{roi}}}{2}\right)}, \quad 
    W_\text{d} = \mathrm{e}^{-\mathrm{j} \frac{2\pi}{L\cdot M}}.
    \label{eq:czt_vel_params}
\end{equation}
The final 2D high-resolution image $\mathbf{Z}_\text{rd}$ is obtained by

\begin{equation}
    \mathbf{Z}_\text{rd}(k, p) = \sum_{m=0}^{M-1} \mathbf{Z}_\text{r}(k, m) A_\text{d}^{m} W_\text{d}^{mp},
    \label{eq:czt_vel_sum}
\end{equation}
where $p \in \{0, \dots, N_\text{czt}-1\}$.

The sub-bin peak location $(k_h^*, p_h^*)$ for target $h$ is identified by maximizing the magnitude of the high-resolution image $|\mathbf{Z}_\text{rd}|^2$. To facilitate the physical parameter mapping, the fractional index offsets $\Delta k_h$ and $\Delta p_h$ are defined relative to the center of the zoomed window as follows

\begin{equation}
    \Delta k_h = \frac{k_h^* - N_\text{czt}/2}{L}, \quad \Delta p_h = \frac{p_h^* - N_\text{czt}/2}{L}.
    \label{eq:offsets}
\end{equation}

\noindent These offsets represent the deviation from the coarse indices $(\hat{n}_h, \hat{m}_h)$ in units of the original bin width. The precise physical delay $\hat{\tau}_{h}$ and the Doppler frequency $\hat{f}_{\text{D,h}}$ are then calculated as

\begin{equation}
    \hat{\tau}_{h} = \frac{\hat{n}_h + \Delta k_h}{B}, \quad
    \hat{f}_{\text{D,h}} = \frac{\hat{m}_h + \Delta p_h}{T},
    \label{eq:fine_params}
\end{equation}

\noindent where $B$ is the signal bandwidth.

Additionally, the optional quadratic interpolation technique described in \cite{Braun_diss} can be applied to further refine the peak estimates obtained from the \ac{CZT}. By approximating the main lobe of the window function as a two-dimensional parabola, fractional corrections to the \ac{CZT} peak indices $(k_h^*, p_h^*)$ can be computed. 
To simplify the notation, let us define the fine-resolution power at indices $(k,p)$ as $S[k,p] = |\mathbf{Z}_\text{rd}[k, p]|^2$. The refined range index $\tilde{k}_h$ and Doppler index $\tilde{p}_h$ are given by \cite{Braun_diss},

\begin{equation}
    \tilde{k}_h = k_h^* + \frac{S(k_h^*-1, p_h^*) - S(k_h^*+1, p_h^*)}{2\left(S(k_h^*-1, p_h^*) + S(k_h^*+1, p_h^*) - 2S(k_h^*, p_h^*)\right)},
    \label{eq:quad_interp_range}
\end{equation}

\begin{equation}
    \tilde{p}_h = p_h^* + \frac{S(k_h^*, p_h^*-1) - S(k_h^*, p_h^*+1)}{2\left(S(k_h^*, p_h^*-1) + S(k_h^*, p_h^*+1) - 2S(k_h^*, p_h^*)\right)},
    \label{eq:quad_interp_doppler}
\end{equation}
to have continuous estimation. However, as it is shown in measurement results, even without this optional process, interference cancellation and weak target detection can be done successfully.
\section{Proposed Interference Mitigation Algorithms}
\label{sec:proposed_methods}
Based on computational complexity difference, we propose two distinct frameworks. Each framework reconstructs target's echo based on delay, Doppler and complex attenuation factor estimates. Before moving on to the details of each algorithm, the calculation of the attenuation factor magnitude, which is used by both methods, is explained. To estimate the complex attenuation factor $\alpha_h$, the signal amplitude is reconstructed by compensating for signal-dependent losses as well as system processing gains. In \eqref{eq:Y(p)}, the useful signal component is initially derived under the assumption of a moderate Doppler shift. More generally, the useful signal power can be expressed for arbitrary Doppler shifts by explicitly accounting for both Doppler-induced attenuation and excess delay. Using the Dirichlet kernel representation, the useful signal power is given by
\begin{equation}
	P_u^h = \underbrace{{\eta_h}^2}_{\text{L}_\text{ISI}(\hat{\tau}_h)} \cdot \underbrace{\left| 
	\frac{
		\sin\left( \dfrac{\pi f_{\text{D},h} \eta_h}{\Delta f} \right)
	}{
		N \cdot \sin\left( \dfrac{\pi f_{\text{D},h} \eta_h}{N \Delta f} \right)
	} 
	\right|^2}_{\text{L}_\text{Dop}(\hat{\tau}_h,\hat{f}_{\text{D},h})} \cdot |\tilde{\alpha}_h|^2.
	\label{eq:Pu}
\end{equation}
For targets that correspond to the delay smaller than the cyclic prefix, $\eta_h = 1$, and the signal is lossless. For excessive delays, $\eta_h$ decreases linearly, introducing an \ac{ISI} power loss factor of $\text{L}_{\text{ISI}} = \eta_h^2$ and a Doppler spreading loss $L_{\text{Dop}}$ characterized by the widened Dirichlet kernel in (\ref{eq:Pu}).

 We utilize the fine-resolution peak power $|\mathbf{Z}_\text{rd}[k_h^*, p_h^*]|^2$ obtained from the \ac{CZT}, which provides a better estimate than the coarse grid peak. The magnitude estimate $|\hat{\alpha}_h|$ is derived by inverting (\ref{eq:Pu})

\begin{equation}
    |\hat{\alpha}_h| = \sqrt{ \frac{|\mathbf{Z}_\text{rd}(k_h^*, p_h^*)|^2}{G_\text{p} \cdot \text{L}_{\text{win}} \cdot \text{L}_{\text{ISI}}(\hat{\tau}_h) \cdot \text{L}_{\text{Dop}}(\hat{\tau}_h, \hat{f}_{\text{D,h}})} },
    \label{eq:alpha_mag}
\end{equation}

\noindent where 
\begin{equation}
    \text{L}_{\text{win}} = \left| \frac{1}{NM} \sum_{n,m} \mathbf{H}_{\text{window}}(n,m) \right|^2,
\end{equation}
which denotes the peak power loss due to the 2D windowing.
Dirichlet kernel term in \eqref{eq:Pu} is the loss related to the Doppler shift but it is a function of $\eta$. This broadening occurs because the effective rectangular window in the time domain shortens as the delay increases, causing the main lobe of the Dirichlet kernel in the frequency domain to widen. Therefore, while estimating the received signal power for each target, we will consider power losses resulting from excessive delay and velocity of the target.
\subsection{JIC-CC: Efficient Cancellation with Coherent Recovery}
The proposed \ac{JIC-CC} framework adopts similar cancellation process to the iterative cancel-then-detect architecture of \ac{SIC} scheme in \cite{Li_SIC}, yet it differs in the interference cancellation and parameter estimation stages. Specifically, the \ac{SIC} method described in \cite{Li_SIC} attempts to estimate the complex attenuation factor $\hat{\alpha}_h$ via a least squares projection, utilizing the estimated delay $\hat{\tau}_h$ and Doppler shift $\hat{f}_{\text{D},h}$. The estimator is given by \cite{Li_SIC,keskin2025}:

\begin{equation}
    \hat{\alpha}_h = \frac{\mathbf{b}^\text{T}(\hat{\tau}_h) \left( \mathbf{Y}_{\text{free}} \odot \mathbf{X}^* \right) \mathbf{c}(\hat{f}_{\text{D},h})}{\|\mathbf{b}(\hat{\tau}_h)\|^2 \|\mathbf{c}(\hat{f}_{\text{D},h})\|^2},
    \label{eq:projection}
\end{equation}

\noindent where $\mathbf{X}^*$ denote the conjugate of transmit frame and $\mathbf{Y}^{\text{free}}_{h} \in \mathbb{C}^{N \times M}$ is the interference-free observation which is also expressed for $m$th symbol in \eqref{eq:expr_Li}. 
The approach in \cite{Li_SIC} attempts to reconstruct the \ac{ISI} and \ac{ICI} terms, $Y^\text{ISI}_h$ and $Y^\text{ICI}_h$, for targets located outside the \ac{ISI}-free region. By iteratively subtracting these components in the frequency domain, the method aims to suppress the interference-noise floor and reveal targets undetectable by conventional processing. However, with this approach, interference effects can be cancelled for on-grid targets but, targets that are below interference-noise floor level due to window mismatch will be missed. Additionally, calculation of the complex attenuation factor $\alpha_h$ via least square estimation can lead to poor estimations when a target is located in off-grid in the radar image. To address these limitations, the proposed \ac{JIC-CC} framework introduces a detection-recovery loop.
Specifically, the \ac{CZT} \cite{CZT_rabiner} is employed to refine the complex amplitude and phase estimates. Furthermore, interference contributions from all detected targets are subtracted jointly as in \eqref{eq:Y_h_in_matrix_form}, and the resulting clean signal $\mathbf{Y}^{c}$ is utilized in frequency-domain coherent compensation to recover the signal energy for targets with excessive delay in the subsequent symbol.

\subsubsection{Algorithm Description}  
Upon detection of targets with conventional processing, the process transitions to the estimation stage. To minimize parameter errors, the \ac{CZT} is utilized to estimate $(\hat{\tau}, \hat{f}_\text{D})$ with super-resolution precision.
In this algorithm, signal cleaning is directly done in the frequency domain. Therefore, estimated phase $\hat{\theta}_{h}$ of the target $h$ can be directly extracted from phase of $\mathbf{Z}_\text{rd}(k_h^*, p_h^*)$.

The complex attenuation factor $\hat{\alpha}_h$ is constructed as 
\begin{equation}
    \hat{\alpha}_h = |\hat{\alpha}_h| \cdot \mathrm{e}^{\mathrm{j} \hat{\theta}_h},
    \label{eq:alpha_final}
\end{equation}
by combining the magnitude estimation in \eqref{eq:alpha_mag} with the phase estimate $\hat{\theta}_h$.

Let $\mathcal{P}_\text{init}$ denote the detected target list with conventional processing. For detected targets with conventional radar processing using \ac{CFAR}, the frequency domain signal for all detected targets in $\mathcal{P}_\text{init}$ can be regenerated based on delay and Doppler estimations $\hat{\tau}_h,\hat{f}_{\text{D},h}$ and those can be removed from received frequency domain signal by 
\begin{equation}
\mathbf{Y}^\text{c} = \mathbf{Y}-\sum\limits_{h \in \mathcal{P}_\text{init}}\mathbf{Y}_h (\hat{\alpha}_h,\hat{\tau}_h,\hat{f}_{\text{D},h}).
\end{equation}
Let $Y_m^\text{c}(k)$ denote the value in $k$th subcarrier and $m$th symbol in $\mathbf{Y}^\text{c}$.  
\ac{FDCC} can be applied to the resulting signal by \cite{geiger_MTCC}
\begin{equation}
	Y_{m}^\text{c-FDCC}(k) = Y^\text{c}_{m}(k) + C(k)  Y^\text{c}_{m+1}(k),
	\label{eq:FDCC} 
\end{equation}
where $C(k) = \mathrm{e}^{-\mathrm{j} 2 \pi k N_\text{cp}/N }$. Due to missing \ac{CP} in frequency domain, full compensation of the signal power loss is not possible. However, with this approach, it is possible to strengthen signal power for targets outside \ac{ISI}-free region, especially if $N_\text{cp}\ll N$. Additionally, since next \ac{OFDM} symbol is fully added to previous symbol in this technique, additional interference is introduced. However, since the dominant echoes from detected targets have already been subtracted from the received signal, this residual interference is minimal, typically falling below the thermal and quantization noise floor. Consequently, conventional radar processing can be repeated by using $Y_{m}^\text{c-FDCC}(k)$ and the weak targets in the environment can be detected.

\begin{algorithm}[!t]
	\caption{JIC-CC with FDCC-Aided Recovery}
	\label{alg:jic_cc}
	\begin{algorithmic}[1]
		\renewcommand{\algorithmicrequire}{\textbf{Input:}}
		\renewcommand{\algorithmicensure}{\textbf{Output:}}
		\REQUIRE Received frame $\mathbf{Y}$, reference $\mathbf{X}$, \ac{CZT} parameters ($B_\text{roi}$, $L$), $\mathbf{H}_\text{window}$
		\ENSURE Reconstructed Image $\mathbf{I}_\text{f}$, Target List $\mathcal{P}_\text{f}$
		\STATE \textbf{Initialization:} $\mathbf{Y}^\text{canc} \leftarrow \mathbf{0}$
		\STATE \textbf{Step 1: Strong Target Detection}
		\STATE $\mathbf{I}_\text{init} \leftarrow \text{ConventionalRadarProcessing}(\mathbf{Y}, \mathbf{X})$
		\STATE $\mathcal{P}_\text{init} \leftarrow \text{CFAR}(\mathbf{I}_\text{init})$
		
		\STATE \textbf{Step 2: Precision Cancellation}
		\FOR{each target $h \in \mathcal{P}_\text{init}$}
		\STATE Estimate $\hat{\tau}_h, \hat{f}_{\text{D},h}, \hat{\theta}_h$ using \textbf{CZT} \COMMENT{\eqref{eq:fine_params}}
		\STATE Estimate complex attenuation factor $\hat{\alpha}_{h}$ \COMMENT{\eqref{eq:alpha_mag}}
		
		\STATE Construct FD Signal $\mathbf{Y}_h(\hat{\alpha}_h,\hat{\tau}_h,\hat{f}_{\text{D},h})$ \COMMENT{\eqref{eq:Y_h_in_matrix_form}}
		\STATE $\mathbf{Y}^\text{canc} \leftarrow \mathbf{Y}^\text{canc} + \mathbf{Y}_h(\hat{\alpha}_h,\hat{\tau}_h,\hat{f}_{\text{D},h})$
		\ENDFOR
		\STATE $\mathbf{Y}^\text{c} \leftarrow \mathbf{Y} - \mathbf{Y}^\text{canc}$
		
		\STATE \textbf{Step 3: FDCC-Aided Weak Target Detection}
		\STATE $\mathbf{Y}^\text{c-FDCC} \leftarrow \text{ApplyFDCC}(\mathbf{Y}^\text{c})$ \COMMENT{\eqref{eq:FDCC}}
		\STATE $\mathbf{I}_\text{clean} \leftarrow \text{ConventionalRadarProcessing}(\mathbf{Y}^\text{c-FDCC}, \mathbf{X})$
		\STATE $\mathcal{P}_\text{weak} \leftarrow \text{CFAR}(\mathbf{I}_\text{clean})$
		\STATE $\mathcal{P}_\text{f} \leftarrow \mathcal{P}_\text{init} \cup \mathcal{P}_\text{weak}$
		
		\STATE \textbf{Step 4: Final Restoration}
		\STATE $\mathbf{I}_{f} \leftarrow \text{RestoreTargetsShape}(\mathbf{I}_\text{clean}, \mathcal{P}_\text{f})$
		\RETURN $\mathbf{I}_\text{f}, \mathcal{P}_\text{f}$
	\end{algorithmic}
\end{algorithm}

\subsection{FR-SW: Full-Reconstruction Sliding Window}

While the \ac{FDCC} approach compensates for power loss in the frequency domain, the \ac{SW} approach addresses the excessive delay problem in the time domain. By exploiting the cyclic nature of the \ac{CP}, we can iteratively shift the observation window to bring distant targets into the ISI-free processing region. To prevent interference-noise floor increase during this process, we employ a hybrid joint interference cancellation and stitching strategy.

\subsubsection{Algorithm Description}
The algorithm first performs standard radar processing on the received time-domain signal $\mathbf{y}[n]$ to identify strong targets. For each detected target $h$, we refine the range and Doppler estimates $(\hat{\tau}_h, \hat{f}_{\text{D},h})$ using the \ac{CZT} method described in Section \ref{sec:signal_model}. The magnitude of complex attenuation factor $|\hat{\alpha}_h|$ is estimated by compensating for the \ac{ISI} and windowing losses as derived in \eqref{eq:alpha_mag}.
 
For the initial phase estimation, it is critical to recognize that the complex phase observed at the range-Doppler peak does not correspond to the target's initial phase at $t=0$.
Instead, it includes a deterministic rotation caused by the Doppler shift up to the effective processing window.
This accumulated rotation is range-dependent, as the effective window shifts for targets subject to delay-induced truncation.
Therefore, to ensure precise coherent cancellation, we apply a phase correction that explicitly compensates for this Doppler-induced rotation, recovering the true phase relative to the global time reference.
Finally, using the fully estimated parameter set $\{\hat{\alpha}_h, \hat{\tau}_h, \hat{f}_{\text{D},h}, \hat{\theta}_h\}$, the received signal for target $h$ is reconstructed and subtracted from the input stream and this process is done for all detected targets jointly. Specifically, the interference cancellation is performed in the time domain by subtracting the reconstructed signal estimates for each target. Based on defined discrete sampling times $t = n T_s$, the clean signal is given by

\begin{equation}
    y_\text{clean}[n] = y[n] - \underbrace{\sum_{h \in \mathcal{P}_\text{init}} \hat{\alpha}_h \left. \left( x(t - \hat{\tau}_h) \mathrm{e}^{\mathrm{j} 2\pi \hat{f}_{\text{D},h} t} \right) \right|_{t=n T_s}}_{y^h[n]},
    \label{eq:td_cancellation}
\end{equation}

\noindent where $x(t)$ denotes the continuous-time transmitted waveform.

After cleaning, the residual signal $\mathbf{y}_\text{clean}$ contains the weak targets potentially located beyond the maximum \ac{ISI}-free range. To recover these, the final high-range image $\mathbf{I}_\text{f}$ is constructed by stitching together "fractional" radar images. 

The algorithm performs $S = \lceil N/N_\text{cp} \rceil$ iterations. In the $s$th iteration (where $s=0, \dots, S-1$), the clean time-domain signal is shifted by $s \cdot N_\text{cp}$ samples:

\begin{equation}
    y^{s}_\text{shift}[n] = y_\text{clean}[n + s \cdot N_\text{cp}].
    \label{eq:time_shift}
\end{equation}

\noindent Standard OFDM radar processing is applied to this shifted signal to obtain an auxiliary range-Doppler image $\mathbf{I}_\text{aux}^{s}$. Due to the time shift, physical targets located at delay $\tau \approx s \cdot T_\text{cp}$ in the original frame now appear near zero delay in $\mathbf{I}_\text{aux}^{s}$, placing them in the high-SNR, ISI-free region.

\begin{algorithm}[!t]
	\caption{SIC-Aided Sliding Window Stitching}
	\label{alg:sliding_window_stitch}
	\begin{algorithmic}[1]
		\renewcommand{\algorithmicrequire}{\textbf{Input:}}
		\renewcommand{\algorithmicensure}{\textbf{Output:}}
		\REQUIRE Received time domain signal $\mathbf{y}$, reference $\mathbf{X}$, \ac{CZT} parameters ($B_\text{roi}$, $L$), $\mathbf{H}_\text{window}$
		\ENSURE Final Image $\mathbf{I}_\text{f}$, Target List $\mathcal{P}_\text{f}$
		
		\STATE \textbf{Step 1: Initial Detection \& Cleaning}
		\STATE $\mathbf{I}_\text{init} \leftarrow \text{ConventionalRadarProcessing}(\mathbf{y}, \mathbf{X})$
		\STATE $\mathcal{P}_\text{init} \leftarrow \text{CFAR}(\mathbf{I}_\text{init})$
		\STATE $\mathbf{y}_\text{clean} \leftarrow \mathbf{y}$, $\mathcal{P}_\text{f} \leftarrow \mathcal{P}_\text{init}$
		
		\FOR{each target $h \in \mathcal{P}_\text{init}$}
			\STATE Estimate $\hat{\tau}_h, \hat{f}_{\text{D},h},\hat{\theta}_h$ via CZT and phase calibration \COMMENT{\eqref{eq:fine_params}}
			\STATE Estimate $\hat{\alpha}_h$ including losses \COMMENT{\eqref{eq:alpha_mag}}
			\STATE Generate $y^h[n]$ and update: $\mathbf{y}_\text{clean} \leftarrow \mathbf{y}_\text{clean} - y^h[n]$
		\ENDFOR
		
		\STATE \textbf{Step 2: Fractional Stitching Loop}
		\STATE Initialize $\mathbf{I}_\text{f}$
		\STATE $S \leftarrow \lceil N / N_\text{cp} \rceil$
		
		\FOR{$s = 0$ \TO $S-1$}
			% --------------------------

			\STATE $\mathbf{y}^s_\text{shift} \leftarrow \text{ShiftSignal}(\mathbf{y}_\text{clean}, s \cdot N_\text{cp})$ \COMMENT{\eqref{eq:time_shift}}
			\STATE $\mathbf{I}_\text{aux}^s \leftarrow \text{FractionalRadarProcessing}(\mathbf{y}^s_\text{shift}, \mathbf{X})$
            \STATE $\mathcal{P}_\text{aux} \leftarrow \text{CFAR}(\mathbf{I}_\text{aux}^s)$
		\STATE $\mathcal{P}_\text{f} \leftarrow \mathcal{P}_\text{f} \cup \mathcal{P}_\text{aux}$
			\STATE \textbf{Stitch:} Extract valid range
            \STATE $R_{\text{start}} \leftarrow s \cdot N_\text{cp}$
            \STATE $L_{\text{seg}} \leftarrow \min(N_\text{cp}, N - R_{\text{start}})$
            \STATE $\mathbf{I}_\text{f}[R_{\text{start}} : R_{\text{start}} + L_{\text{seg}}-1, :] \leftarrow \mathbf{I}_\text{aux}^s[0 : L_{\text{seg}}-1, :]$
			%\STATE Define range indices: $r \in [0, N_\text{cp}-1]$
			%\STATE Define mapping: $r_\text{act} \leftarrow r + s \cdot N_\text{cp}$
			%\STATE $\mathbf{I}_\text{f}[r_\text{act}, :] \leftarrow \mathbf{I}_\text{aux}[r, :]$
		\ENDFOR
		
		\STATE \textbf{Step 3: Restoration}
		\STATE $\mathbf{I}_\text{f} \leftarrow \text{RestoreTargets}(\mathbf{I}_\text{f}, \mathcal{P}_\text{f})$
		\RETURN $\mathbf{I}_\text{f}, \mathcal{P}_\text{f}$
	\end{algorithmic}
\end{algorithm}

The final image is constructed by extracting the valid \ac{ISI}-free region from each auxiliary image and mapping it to the corresponding absolute delay indices in the final image. To handle the general case where the total number of subcarriers $N$ is not an integer multiple of $N_\text{cp}$, we define the length of the valid segment for the $s$th window shift as
\begin{equation}
    L_\text{seg} = \min(N_\text{cp}, N - s \cdot N_\text{cp}).
    \label{eq:segment_len}
\end{equation}
The stitching operation is then performed as
\begin{equation}
    \mathbf{I}_\text{f}[r + s \cdot N_\text{cp}, m] = \mathbf{I}_\text{aux}^{s}[r, m],
    \label{eq:stitching}
\end{equation}
valid for local range indices $r \in [0, L_\text{seg}-1]$. This process effectively solves the range ambiguity by stitching together locally valid segments of length $L_\text{seg}$ to form a complete, extended radar image up to the unambiguous range. Finally, the strong targets removed in the first step are restored to $\mathbf{I}_\text{f}$ to complete the scene.
\section{Simulation Results and Complexity Analysis}
\label{sec:sim_res}
\begin{figure}
    \centering
    \psfrag{Target 2 RCS (dBsm)}[c][c]{\scriptsize Target 1 RCS (dBsm)}
    \psfrag{SINR Image (dB)}[c][c]{\scriptsize Mean SINR Image (dB)}
    \psfrag{0}[c][c]{\scriptsize $0$}
    \psfrag{10}[c][c]{\scriptsize $10$}
    \psfrag{20}[c][c]{\scriptsize $20$}
    \psfrag{30}[c][c]{\scriptsize $30$}
    \psfrag{40}[c][c]{\scriptsize $40$}
    \psfrag{50}[c][c]{\scriptsize $50$}
    \psfrag{60}[c][c]{\scriptsize $60$}
    \psfrag{0}[c][c]{\scriptsize $0$}
    \psfrag{5}[c][c]{\scriptsize $5$}
    \psfrag{10}[c][c]{\scriptsize $10$}
    \psfrag{15}[c][c]{\scriptsize $15$}
    \psfrag{20}[c][c]{\scriptsize $20$}
    \psfrag{-5}[c][c]{\scriptsize $-5$}
    \psfrag{-10}[c][c]{\scriptsize $-10$}
    \psfrag{-15}[c][c]{\scriptsize $-15$}
    \includegraphics[width=0.8\linewidth]{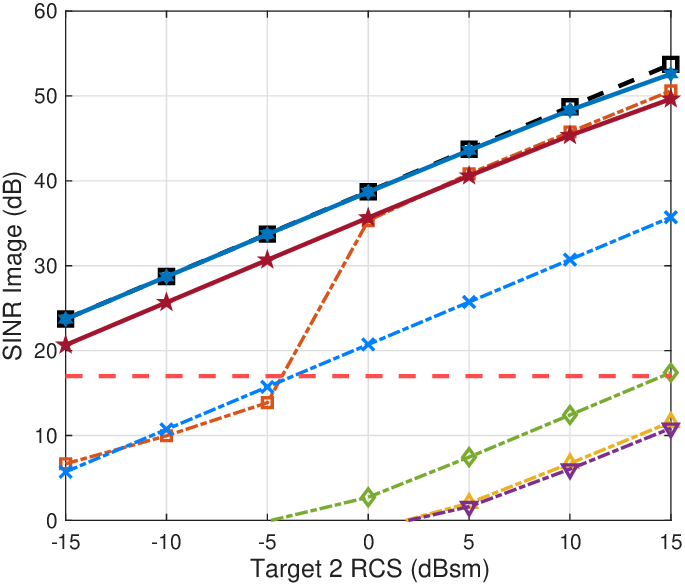}
\caption{SINR comparison for the weak target against RCS. 
    The empirical detection threshold is indicated by 
    ({\color[rgb]{1,0,0}\textbf{\rule[0.5ex]{0.5em}{1.25pt}\hspace{0.2em}\rule[0.5ex]{0.5em}{1.25pt}}}). 
    The theoretical ideal SNR is shown in 
    ({\color[rgb]{0,0,0}\textbf{\rule[0.5ex]{0.4em}{1.25pt}\hspace{0.1em}\raisebox{-0ex}{\tiny $\square$}\hspace{0.1em}\rule[0.5ex]{0.4em}{1.25pt}}}). 
    %Also conventional radar processing \ac{SINR} image is indicated by ({\color[rgb]{0.5,0.5,0.5}\textbf{\rule[0.5ex]{0.6em}{1.25pt}\hspace{0.1em}\raisebox{-0ex}{\tiny $\circ$}\hspace{0.1em}\rule[0.5ex]{0.4em}{1.25pt}}})
    The existing methods are shown with lines: 
    SW ({\color[rgb]{0,0.5,1}\textbf{\rule[0.5ex]{0.6em}{1.25pt}\hspace{0.1em}\raisebox{-0ex}{\tiny $\times$}\hspace{0.1em}\rule[0.5ex]{0.4em}{1.25pt}}}), 
    15 iteration SIC 
    ({\color[rgb]{0.85,0.33,0.10}\textbf{\rule[0.5ex]{0.6em}{1.25pt}\hspace{0.1em}\raisebox{-0ex}{\tiny $\square$}\hspace{0.1em}\rule[0.5ex]{0.4em}{1.25pt}}}), 
    TDCC 
    ({\color[rgb]{0.93,0.69,0.13}\textbf{\rule[0.5ex]{0.6em}{1.25pt}\hspace{0.1em}\raisebox{0ex}{\tiny $\triangle$}\hspace{0.1em}\rule[0.5ex]{0.4em}{1.25pt}}}), 
    FDCC 
    ({\color[rgb]{0.49,0.18,0.56}\textbf{\rule[0.5ex]{0.6em}{1.25pt}\hspace{0.1em}\raisebox{0ex}{\scriptsize $\triangledown$}\hspace{0.1em}\rule[0.5ex]{0.4em}{1.25pt}}}), 
    and MTCC 
    ({\color[rgb]{0.47,0.67,0.19}\textbf{\rule[0.5ex]{0.6em}{1.25pt}\hspace{0.1em}\raisebox{-0ex}{\tiny $\Diamond$}\hspace{0.1em}\rule[0.5ex]{0.4em}{1.25pt}}}). 
    The proposed methods are shown with: 
    SIC-CC 
    ({\color[rgb]{0.64,0.08,0.18}\textbf{\rule[0.5ex]{1.2em}{1.25pt}\hspace{-1em}\raisebox{0.1ex}{\scriptsize $\bigstar$}}}) 
    and FR-SW 
    ({\color[rgb]{0,0.45,0.74}\textbf{\rule[0.5ex]{1.2em}{1.25pt}\hspace{-1em}\raisebox{-0.1ex}{\scriptsize $\bigstar$}}}).
    }
    \label{fig:sim_comparison_with_literature}
\end{figure}

To benchmark the proposed algorithms against the existing literature, the detection performance for weak targets near the maximum unambiguous range is evaluated. The simulation parameters are listed in Table~\ref{tab:sim_pars}. To ensure a fair comparison, windowing is omitted for all schemes and and targets are ideally placed on the range-Doppler grid, i.e., integer multiples of range and velocity resolution. 
%\commentsm{I would have preferred windowing everywhere}
Consistent with the previous analysis, a worst-case scenario is adopted wherein a strong interfering target ($20~\text{dBsm}$) is positioned at the range yielding maximum interference power, as illustrated in Fig.~\ref{fig:motivation}. A secondary weak target is placed near the unambiguous range limit where $R_\text{max,unamb} = 4914$\,m. To establish a fair baseline comparison isolated from spectral leakage effects, this target is intentionally placed exactly on the range grid at $4839$\,m. The \ac{RCS} of this weak target is swept, and the resulting image \ac{SINR} levels are measured for all considered algorithms. Additionally, a detection threshold of \qty{17}{dB} is indicated in Fig.~\ref{fig:sim_comparison_with_literature} to define the bound for reliable target detection \cite{mandelli_ISAC}.
%\commentsm{Why 4839, are you placing it on-grid? As I guess so, that's why windowing is not so necessary for you, but it's a clear attacking point}

\begin{figure}
    \centering
    \psfrag{Amplitude MAE (dB)}[c][c]{\scriptsize Amplitude MAE (dB)}
    \psfrag{Range offset (m)}[c][c]{\scriptsize Range offset (m)}
    \psfrag{Interference-noise floor (dBm)}[c][c]{\scriptsize Interference-noise floor (dBm)}
    \psfrag{Phase MAE (deg)}[c][c]{\scriptsize Phase MAE $(^\circ)$}
    \psfrag{5}[c][c]{\scriptsize $5$}
   \psfrag{4}[c][c]{\scriptsize $4$}  
   \psfrag{3}[c][c]{\scriptsize $3$}
   \psfrag{2}[c][c]{\scriptsize $2$}  
  \psfrag{1}[c][c]{\scriptsize $1$}
   \psfrag{0}[c][c]{\scriptsize $0$} 
 \psfrag{0.2}[c][c]{\scriptsize $0.2$}
   \psfrag{0.4}[c][c]{\scriptsize $0.4$} 
    \psfrag{-0.2}[c][c]{\scriptsize -$0.2$}
   \psfrag{-0.4}[c][c]{\scriptsize -$0.4$}
    \psfrag{0.1}[c][c]{\scriptsize $0.1$}
   \psfrag{0.2}[c][c]{\scriptsize $0.2$}
   \psfrag{0.3}[c][c]{\scriptsize $0.3$}
   \psfrag{0.4}[c][c]{\scriptsize $0.4$}
   \psfrag{0.5}[c][c]{\scriptsize $0.5$}
  \psfrag{0.6}[c][c]{\scriptsize $0.6$}
 \psfrag{0.8}[c][c]{\scriptsize $0.8$}
   \psfrag{-55}[c][c]{\scriptsize -$55$}
    \psfrag{-60}[c][c]{\scriptsize -$60$}
     \psfrag{-65}[c][c]{\scriptsize -$65$}
      \psfrag{-70}[c][c]{\scriptsize -$70$}
      \psfrag{-75}[c][c]{\scriptsize -$75$}
      \psfrag{-80}[c][c]{\scriptsize -$80$}
    \includegraphics[width=1\linewidth]{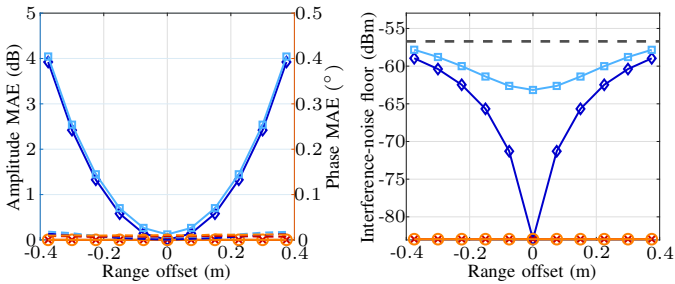}
\caption{Performance analysis of parameter estimation and interference cancellation under on-grid and off-grid conditions. 
    (a) Estimation accuracy: Solid lines represent amplitude \ac{MAE} (left axis), and dashed lines represent phase \ac{MAE} (right axis). 
    (b) Residual interference-noise floor in comparison to the $-82.96$ dBm thermal limit. 
    Initial noise-interference floor with conventional processing is indicated by ({\color[rgb]{0,0,0}\textbf{\rule[0.5ex]{0.6em}{1.25pt}\hspace{0.1em}\rule[0.5ex]{0.4em}{1.25pt}}}). Results are shown for the projection-based estimator \eqref{eq:projection} at $f_\text{D} = 0$ 
    ({\color[rgb]{0,0,0.8}\textbf{\rule[0.5ex]{0.6em}{1.25pt}\hspace{0em}\raisebox{-0ex}{\tiny $\Diamond$}\hspace{0em}\rule[0.5ex]{0.4em}{1.25pt}}}) 
    and $f_\text{D} = 0.1 \Delta f$ 
    ({\color[rgb]{0.3,0.7,1}\textbf{\rule[0.5ex]{0.6em}{1.25pt}\hspace{0em}\raisebox{-0ex}{\tiny $\square$}\hspace{0em}\rule[0.5ex]{0.4em}{1.25pt}}}); 
    and for the \ac{CZT}-based estimator at $f_\text{D} = 0$ 
    ({\color[rgb]{0.8,0,0}\textbf{\rule[0.5ex]{0.6em}{1.25pt}\hspace{0em}\raisebox{-0ex}{\tiny $\times$}\hspace{0em}\rule[0.5ex]{0.4em}{1.25pt}}}) 
    and $0.1 \Delta f$ 
    ({\color[rgb]{1,0.5,0}\textbf{\rule[0.5ex]{0.6em}{1.25pt}\hspace{0em}\raisebox{-0ex}{\tiny $\circ$}\hspace{0em}\rule[0.5ex]{0.4em}{1.25pt}}}).}
    \label{fig:alpha_phase_est}
\end{figure}

As illustrated in Fig.~\ref{fig:sim_comparison_with_literature}, the proposed \ac{FR-SW} algorithm achieves an \ac{SINR} profile that closely matches the theoretical ideal. This demonstrates the efficiency of the sliding window technique combined with strong target cancellation in fully recovering weak targets. Conversely, the proposed \ac{JIC-CC} algorithm exhibits slightly suboptimal performance compared to the ideal curve. This deviation can be attributed to the signal power loss inherent to the \ac{FDCC} operation when the \ac{CP} is absent and thermal noise power increase during this process. In the high \ac{RCS} regime ($>-5$ dBsm), the \ac{SIC} algorithm from \cite{Li_SIC} outperforms slightly \ac{JIC-CC} but remains inferior to \ac{FR-SW}. This behavior is expected; while the algorithm used 15 iterations, further increasing this number could theoretically allow convergence to the ideal bound, but at the cost of prohibitive computational complexity. A more critical limitation of \ac{SIC} is observed at low \ac{RCS} values ($\le -10$ dBsm). As shown, a sharp performance discontinuity occurs where the target \ac{SINR} falls below the detection threshold (17 dB). In this region, the algorithm fails to reliably detect and reconstruct the signal $\mathbf{Y}_{\text{free}}$, leading to a breakdown in cancellation performance. 

Additionally, \ac{SW} algorithm proposed in \cite{Xu_SW} is also investigated. As explained earlier, targets inside \ac{ISI}-free region are detected at every iteration and their contribution to received signal is removed iteratively and overall radar image is constructed by stitching individual radar processing results up to $R_\text{max,ISI}$ at every iteration in \cite{Xu_SW}. However, this technique leads to high interference-noise floor in earlier iterations until the strong target that is outside \ac{ISI}-free region is detected and removed. Therefore, mean \ac{SINR} image reduction is also observed with that technique as well. 

The investigation also evaluates the \ac{MTCC} algorithm, which is proposed in \cite{geiger_MTCC}. While \ac{MTCC} improves upon standard techniques like \ac{FDCC} and \ac{TDCC} by applying thresholding in multi-target scenarios, it still suffers from \ac{SINR} loss. When a strong target lies outside the \ac{ISI}-free region, it encounters \ac{ISI} and \ac{ICI} that elevates the interference-noise floor. Since \ac{MTCC} suppresses only the peak power of strong targets via thresholding, the residual interference (e.g., sidelobes) remains, resulting in the observed degradation.

In practical scenarios, off-grid targets introduce spectral leakage that can further degrade performance. Consequently, this section investigates the necessity of \ac{CZT}-based parameter estimation to address these off-grid effects.
This investigation considers a single strong target ($20~\text{dBsm}$) located outside the \ac{ISI}-free region. Given a noise figure of \mbox{$\text{NF}=8~\text{dB}$}, the theoretical thermal noise floor following ideal cancellation is $-82.96~\text{dBm}$.
To evaluate robustness against off-grid effects, the target range is swept across the interval \mbox{$[R-\Delta R/2, R+\Delta R/2]$} centered at $R=750~\text{m}$. We analyze the \ac{MAE} of the estimated attenuation factor $|\hat{\alpha}_h|$ in dB scale and phase $\hat{\theta}_h$ for both the projection-based technique (via \eqref{eq:projection}) with iterative cleaning process described in \cite{Li_SIC} and the \ac{CZT}-based estimation described in this paper.

The estimation and cancellation performances are presented in Fig.~\ref{fig:alpha_phase_est}. As shown in Fig.~\ref{fig:alpha_phase_est}(a), the \ac{CZT}-based method achieves near-perfect estimation regardless of grid alignment. Conversely, the projection-based estimator exhibits higher amplitude \ac{MAE}, which degrades significantly with velocity and off-grid offsets. Additionally, both estimation techniques demonstrate high precision in terms of phase \ac{MAE} with errors remaining below $0.05^\circ$ for the considered system configuration and scenarios. Fig.~\ref{fig:alpha_phase_est}(b) demonstrates that these discrepancies, combined with delay and Doppler estimation errors inherent to off-grid scenarios in the projection-based technique, translate into significant residual interference. Specifically, the projection approach fails to reach the thermal noise floor for any scenario other than static, on-grid targets.

Furthermore, while the results in Fig.~\ref{fig:alpha_phase_est} omit the windowing process, the application of a practical window function to suppress sidelobes would lead to an offset in the attenuation factor estimation via the projection-based approach. Consequently, any applied windowing function leads to a substantial residual interference floor even for on-grid targets. In contrast, the proposed frameworks inherently mitigate this offset by explicitly compensating for the 2D windowing loss factor $L_\text{win}$, as derived in \eqref{eq:alpha_mag}, ensuring robust cancellation under fully practical conditions.
\begin{table}
    \vspace{-0.2cm}
    % --- Style Definitions from Table 2 ---
    \renewcommand{\arraystretch}{1.5}
    \setlength{\arrayrulewidth}{.1mm}
    \setlength{\tabcolsep}{4pt}
    \centering
    \captionsetup{width=43pc,justification=centering,labelsep=newline}
    
    % --- Content from Table 1 ---
    \caption{\textsc{Computational Complexity Comparison of Algorithms}}
    \label{tab:complexity_comparison}
    
    \begin{tabular}{lll}
        \toprule
        \textbf{Algorithm} & \textbf{Computational Complexity} \\
        \midrule
        TDCC / FDCC \cite{wang_CC} & $\mathcal{O}(MN \log(MN))$ \\
        
        MTCC \cite{geiger_MTCC} & $\mathcal{O}(3 MN \log(MN))$ \\
        
        SIC \cite{Li_SIC} & $\mathcal{O}(N_{\text{iter}} (MN \log(MN) + HMN))$ \\
        
        \ac{SW} \cite{Xu_SW} & $\mathcal{O}((S+1) MN \log(MN) + H M(N+N_{\text{cp}}))$ \\
        
        \textbf{Proposed JIC-CC} & $\mathcal{O}(2 MN \log(MN) + H_\text{d} (C_{\text{czt}} + MN))$ \\
        
        %\textbf{Proposed FR-SW} & $\mathcal{O}(S MN \log(MN) + H_\text{d} (C_{\text{czt}} + C_\text{tdr}))$ \\
        \textbf{Proposed FR-SW} & $\mathcal{O}((S+1) MN \log(MN) + H_\text{d} (C_{\text{czt}} + C_\text{tdr}))$ \\
        \bottomrule
    \end{tabular}
    
    % --- Footer Note (kept from Table 1, formatted to fit) ---
    \vspace{1ex}
    \raggedright
    \footnotesize{\textit{Note:} The term $H_\text{d}$ accounts for number of detected targets with initial radar processing.}
  \vspace{-0.2cm}  
\end{table}

Additionally, the computational costs of the proposed algorithms are compared against state-of-the-art techniques in Table~\ref{tab:complexity_comparison}. The analysis considers the three primary processing stages:
(i) the radar processing with 2D-\ac{FFT} ($\mathcal{O}(MN \log(MN))$);
(ii) the parameter estimation cost (iterative or \ac{CZT}-based);
(iii) the interference reconstruction cost.

A critical distinction lies in the domain of reconstruction. Time-domain approaches, such as the sliding window methods (\ac{SW} \cite{Xu_SW} and the proposed \ac{FR-SW}), must generate the full signal vector including the cyclic prefix to effectively mitigate the interference. 
This reconstruction cost is denoted as $C_\text{tdr}$ in Table~\ref{tab:complexity_comparison}.
In our simulations and measurements, $C_\text{tdr}$ is modeled based on \ac{FFT} operations to ensure exact fractional delay and Doppler calculation, scaling as \mbox{$\mathcal{O}(M(N+N_{\text{cp}}) \log(M(N+N_{\text{cp}})))$},
while real-time hardware could reduce this to linear complexity $\mathcal{O}(K \cdot M(N+N_{\text{cp}}))$ using $K$-tap polyphase \ac{FIR} filters. Since the receiver must iteratively transform the signal between domains for every shift $S = \lceil N / N_\text{cp} \rceil$, sliding window techniques result in a multiplicative scaling of the complexity. For the proposed \ac{FR-SW}, the complexity is detailed in Table~\ref{tab:complexity_comparison} for standard 2D \ac{FFT} implementation which is utilized for every sliding window shift and would result in substantial redundancy, as the full spectrum is re-calculated despite only a fraction being relevant to the current window. Therefore complexity can be reduced by employing an output-pruned \ac{FFT} algorithm \cite{Sorensen_PrunedFFT}.
Since each iteration of the \ac{FR-SW} framework isolates a specific frequency sub-band of size roughly $N/S$, the transform can be restricted to evaluate only these required coefficients.
As derived in \cite{Sorensen_PrunedFFT}, computing a $N/S$ outputs from $N$ inputs reduces the computational cost from $\mathcal{O}(N \log N)$ to approximately $\mathcal{O}(N \log (N/S))$.
Consequently, by targeting only the relevant sub-band indices, the complexity for the iterative stage can be effectively reduced by using pruned \ac{FFT} in the sliding window architecture.

In contrast, the proposed \ac{JIC-CC} algorithm operates entirely in the frequency domain, requiring only two 2D \ac{FFT} operations alongside the local \ac{CZT} estimation. By eliminating the multiplicative factor $S$ inherent to sliding window approaches, \ac{JIC-CC} achieves the efficient complexity profile shown in Table~\ref{tab:complexity_comparison}, offering robust performance with a computational burden comparable to existing state-of-the-art algorithms.

Additionally, both proposed algorithms utilize the \ac{CZT} to improve parameter estimation and this cost is denoted as $C_\text{czt}$. The refinement cost for a single target is determined by the 2D Chirp-Z Transform, implemented via Bluestein's \ac{FFT}-based convolution algorithm \cite{bluestein1970}. 
Assuming a single refinement window size $N_{\text{czt}}$ that is negligible compared to the full signal dimensions (i.e., $N_{\text{czt}} \ll \{N, M\}$), the complexity of \ac{CZT} can be expressed as \mbox{$C_{\text{czt}} \approx \mathcal{O}(MN \log N + N_{\text{czt}} M \log M)$}. 

In addition to computational complexity, the evaluated algorithms impose distinct memory and hardware buffering requirements. Frequency-domain approaches, such as the proposed \ac{JIC-CC}, as well as existing methods like \ac{SIC} \cite{Li_SIC} and MTCC \cite{geiger_MTCC}, operate predominantly on the demodulated signal. Consequently, the hardware only needs to buffer the standard $N \times M$ received symbol matrix after \ac{CP} removal, identical to conventional \ac{OFDM} radar processing, alongside the $N \times M$ reference frame. 

In contrast, time-domain interference cancellation and iterative window shifting methods, such as the proposed \ac{FR-SW} and the \ac{SW} algorithm in \cite{Xu_SW}, require buffering the raw, time-domain baseband signal. Specifically, these techniques necessitate buffering $(M+1)(N+N_\text{cp})$ received samples to accommodate the sliding window shifts, alongside $M(N+N_\text{cp})$ samples of the ideal transmit signal to apply precise delay, Doppler, and attenuation corrections during the time-domain removal process. For scenarios requiring large frame sizes, this extensive time-domain buffering can increase the on-chip memory consumption compared to frequency-domain approaches. Furthermore, while \ac{JIC-CC} and \ac{FR-SW} utilize a \ac{CZT} factor ($L$) for precise parameter estimation, the resulting high-resolution image is strictly transient. The system simply extracts the scalar parameters (delay, Doppler, and attenuation) for the detected targets and discards the interpolated grid. Thus, the memory overhead for the high precision parameter estimation via \ac{CZT} adds a negligible burden.

\section{Verification Measurements}
\label{sec:valid}
\begin{table}
\vspace{-0.2cm}
    \renewcommand{\arraystretch}{1.5}
      		
    \centering
\captionsetup{width=43pc,justification=centering,labelsep=newline}
    \caption{\textsc{Considered Measurement Parameters}}
    \label{tab:OFDM_par}
    \begin{tabular}{lll}
        \toprule
        \textbf{Parameters} & \textbf{Value} \\
        \midrule
        Carrier frequency  ($f_\mathrm{c}$) & \qty{3.68}{\giga\hertz} \\
        Frequency bandwidth ($B$) & \qty{500}{\mega\hertz} \\
        Number of subcarriers ($N$) & \num{1024} \\
        Cyclic prefix length ($N_{\mathrm{cp}}$) & \num{256} \\
        OFDM symbols per frame ($M$) & \num{1024}  \\
        Number of targets ($H$) & \num{6}  \\
        Ranges (m) & \num{72}, \num{150}, \num{162}, \num{222}, \num{228}, \num{240} \\
        Velocities (km/h) & \num{0}, \num{-220}, \num{220}, \num{0}, \num{0}, \num{0} \\
        Attenuation factors (dB) & \num{0}, \num{0}, \num{0}, \num{50}, \num{50}, \num{0} \\
        \ac{CZT} factor ($L$) & \num{100} \\
        \bottomrule
    \end{tabular}
    \vspace{-0.2cm}
    \label{tab:meas_param}
\end{table}  

\begin{figure}
    \centering
    \includegraphics[width=0.75\linewidth]{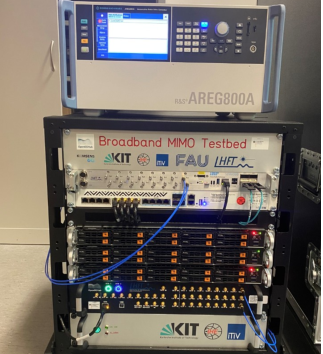}
    \caption{Measurement setup with \ac{ISAC} testbed and radar target simulator.}
    \label{fig:meas_setup}
    \vspace{-0.2cm}
\end{figure}

\begin{figure*}
    \centering
    % colorbar
    \psfrag{-55}[c][c]{\scriptsize -$55$}
    \psfrag{-50}[c][c]{\scriptsize -$50$}
    \psfrag{-45}[c][c]{\scriptsize -$45$}
    \psfrag{-40}[c][c]{\scriptsize -$40$}
    \psfrag{-35}[c][c]{\scriptsize -$35$}
    \psfrag{-30}[c][c]{\scriptsize -$30$}
    \psfrag{-25}[c][c]{\scriptsize -$25$}
    \psfrag{-20}[c][c]{\scriptsize -$20$}
    \psfrag{-15}[c][c]{\scriptsize -$15$}
    \psfrag{-10}[c][c]{\scriptsize -$10$}
    \psfrag{-5}[c][c]{\scriptsize -$5$}
    \psfrag{0}[c][c]{\scriptsize $0$}
    % z-axıs
    \psfrag{-100}[c][c]{\scriptsize -$100$}
    \psfrag{-80}[c][c]{\scriptsize -$80$}
    \psfrag{-60}[c][c]{\scriptsize -$60$}
    \psfrag{-40}[c][c]{\scriptsize -$40$}
    \psfrag{-20}[c][c]{\scriptsize -$20$}
    \psfrag{-0}[c][c]{\scriptsize $0$}
    \psfrag{20}[c][c]{\scriptsize $20$}
    % Velocıty axis ticks
    \psfrag{-500}[c][c]{\scriptsize -$500$}
    \psfrag{0}[c][c]{\scriptsize $0$}
    \psfrag{500}[c][c]{\scriptsize $500$}
    % Range axis ticks
    \psfrag{0}[c][c]{\scriptsize $0$}
    \psfrag{50}[c][c]{\scriptsize $50$}
    \psfrag{100}[c][c]{\scriptsize $100$}
    \psfrag{150}[c][c]{\scriptsize $150$}
    \psfrag{200}[c][c]{\scriptsize $200$}
    \psfrag{250}[c][c]{\scriptsize $250$}
    \psfrag{300}[c][c]{\scriptsize $300$}
    %\psfrag{ZZZ3}[c][c]{\s Normalized Power (dB)}
    \psfrag{Velocity (m/s)}[c][c]{\scriptsize Velocity (m/s)}
    \psfrag{Range (m)}[c][c]{\scriptsize Range (m)}
    \psfrag{Power (dB)}[c][c]{\scriptsize Power (dB)}  

    \includegraphics[width=1\linewidth]{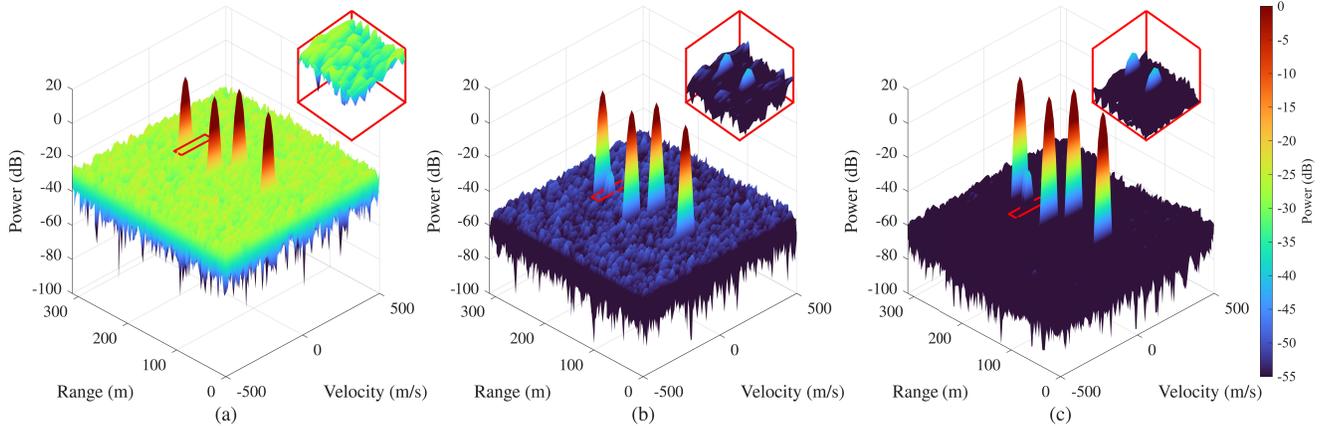}
    \caption{Measurement results with (a) conventional radar processing, \ac{CZT}-based (b) \ac{JIC-CC}, and (c) \ac{FR-SW}.} 
    \label{fig:meas_res}
\end{figure*}

To validate the effectiveness of the proposed \ac{JIC-CC} and \ac{FR-SW} frameworks in a practical environment, proof-of-concept measurements were conducted using a hardware-in-the-loop setup. The measurement setup, illustrated in Fig.~\ref{fig:meas_setup}, comprises a broadband OFDM-based \ac{ISAC} testbed and a Rohde \& Schwarz AREG800A radar target emulator.
Further details regarding the ISAC testbed can be found in \cite{nuss2024testbed} and \cite{karle2024modular}.

The AREG800A is utilized to emulate multiple radar targets with precise range, velocity, and \ac{RCS} characteristics directly in the \ac{RF} domain, providing a controllable and repeatable environment for interference analysis. In this setup, a single baseband module is employed to generate the transmit signal and process the received echo, ensuring intrinsic synchronization between the transmitter and receiver. The system operates at a carrier frequency of \qty{3.68}{\giga\hertz} with a bandwidth of \qty{500}{\mega\hertz}. The specific system parameters configured for this measurement campaign are detailed in Table~\ref{tab:meas_param}. A critical aspect of this experimental design is the specific configuration of the target scenario relative to the \ac{CP} duration. As shown in Table~\ref{tab:meas_param}, the cyclic prefix length is set to $N_\text{cp} = 256$ samples. Given the system bandwidth, this corresponds to a maximum \ac{ISI}-free range of approximately $R_\text{max,ISI}=76.8~\text{m}$. 

To create the severe interference conditions that necessitate the proposed cancellation algorithms, the AREG800A was programmed to generate a multi-target environment where the majority of targets fall outside \ac{ISI}-free range. 
Specifically, the first target is positioned at \qty{72}{\meter}, which lies within the \ac{ISI}-free zone. However, subsequent targets are located at ranges such as \qty{150}{\meter}, \qty{162}{\meter}, and \qty{240}{\meter}, all of which significantly exceed $R_\text{max,ISI}$. These distant targets induce substantial \ac{ISI} and \ac{ICI}, which spill over into the processing window of the first target and elevate the overall interference-noise floor in the radar image. Furthermore, to evaluate the robustness of algorithms in high-dynamic-range scenarios, the emulation includes targets with different attenuation factors (e.g., \qty{0}{\decibel} versus \qty{50}{\decibel}). This configuration creates a challenging masking scenario where strong distant targets (0 dB attenuation) obscure weaker targets (50 dB attenuation), allowing for a rigorous benchmark of the cancellation accuracy and dynamic range improvement offered by the \ac{JIC-CC} and \ac{FR-SW} methods compared to conventional processing.

Furthermore, to evaluate the robustness of algorithms in high-dynamic-range scenarios, the emulation parameters were defined in terms of attenuation factor rather than \ac{RCS}.
This approach allows for the direct control of the received signal power independent of the emulated range, effectively decoupling the \ac{SNR} of targets from the range-dependent free-space path loss.
Consequently, interference power can be controlled, and challenging masking scenarios can be created easily, where distant targets can maintain high interference power.
\begin{table}
\vspace{-0.2cm}
    \renewcommand{\arraystretch}{1.5}      		
    \centering    \captionsetup{width=43pc,justification=centering,labelsep=newline}
    \caption{\textsc{Measured \ac{SINR} image for proposed schemes}}   
    \label{tab:meas_res}
    \begin{tabular}{lll}
        \hline
        \textbf{Processing Method} & \textbf{Image SINR (dB)} \\
        \hline
        Conventional Processing & $<0$ \\
        \textbf{Proposed JIC-CC} & 18.6 \\
        \textbf{Proposed FR-SW} & 25.5 \\
        \hline
    \end{tabular}
    \vspace{-0.2cm}
\end{table}  

The performance of the proposed interference cancellation schemes was validated using a measurement setup involving six targets with ranges and velocities indicated in Table~\ref{tab:meas_param}. Fig.~\ref{fig:meas_res} illustrates the Range-Doppler maps obtained from the conventional processing and the proposed algorithms, while Table~\ref{tab:meas_res} quantitatively summarizes the final \ac{SINR} of the recovered weak targets. As expected, both the \ac{JIC-CC} and the \ac{FR-SW} demonstrate a significant capability to suppress interference, resulting in a notably reduced interference-noise floor compared to the conventional processing baseline. 

A closer inspection of Table~\ref{tab:meas_res} reveals that the \ac{FR-SW} approach achieves a superior interference mitigation performance, yielding a cleaner spectrum and a higher final \ac{SINR} image compared to the \ac{JIC-CC} method. This performance gap is driven by three inherent characteristics of the frequency-domain operation utilized in \ac{JIC-CC}. First, \ac{FDCC} sums adjacent received symbols to recover signal energy, which statistically doubles the thermal noise power, inherently raising the noise floor by approximately 3 dB. Second, because the \ac{CP} is discarded prior to the \ac{FFT}, the frequency-domain compensation cannot recover the fraction of signal energy that fell within the \ac{CP} window, leading to a slight signal power loss. Finally, for environments with high-velocity targets, the pure frequency-domain cancellation does not fully resolve range-Doppler coupling effects, which can leave minor residual interference after cancellation.

However, it is crucial to emphasize that this \ac{SINR} gap narrows significantly in many practical deployments. For scenarios dominated by static or low-velocity interferers, the range-Doppler coupling mismatch becomes negligible. Furthermore, in standard \ac{OFDM} waveforms where the \ac{CP} overhead is kept small to maximize spectral efficiency (i.e., $N_{cp} \ll N$), the unrecoverable power loss in the \ac{CP} window becomes small. Consequently, while \ac{FR-SW} provides better dynamic range and theoretical accuracy, the \ac{JIC-CC} framework offers a highly favorable trade-off. It provides robust interference suppression and reliable weak target detection with substantially lower computational complexity, making it a highly attractive solution for real-time or resource-constrained \ac{ISAC} applications.

\section{Conclusion}
\label{sec:conc}
This paper addressed the challenge of extending the sensing range beyond \ac{ISI}-free range up in \ac{OFDM}-based \ac{ISAC} systems, particularly under high-dynamic-range conditions and realistic scenarios such as off-grid targets. Two algorithms were proposed: \ac{JIC-CC}, which integrates \ac{FDCC} with \ac{CZT}-based estimation for efficient interference cancellation, and \ac{FR-SW}, which shifts the receive window to maximize signal energy and suppress residual interference.

Simulation and measurement results demonstrate that both methods outperform existing algorithms in multi-target scenarios. While \ac{FR-SW} offers the highest dynamic range, \ac{JIC-CC} achieves comparable performance with lower complexity, making it suitable for real-time applications. Future work will explore extensions of the algorithms to \ac{MIMO} systems and extended target models.

\section*{Acknowledgments}
{\linespread{1.1}\selectfont
The authors would like to thank Rohde \& Schwarz for providing the AREG800A automotive radar echo generator used in the measurement campaign.\par}

\bibliographystyle{IEEEtran}
\bibliography{references}

@INPROCEEDINGS{nuss2024testbed,	
	author={B. Nuss and others},	
	booktitle={in Proc. European Wireless},	
	title={Flexible and Scalable Broadband Massive {MIMO} Testbed for Joint Communication and Sensing Applications}, 
	year={2024},
	month={Sept.}
}

@book{Richards2005,
  title={Fundamentals of Radar Signal Processing},
  author={Mark A. Richards},
  year={2005},
  publisher={McGraw-Hill},
  url={https://api.semanticscholar.org/CorpusID:56692486}
}

@techreport{3gpp_ran109_6g_waveform,
  author      = {{3GPP}},
  title       = {{Study on 6G Radio}},
  institution = {3rd Generation Partnership Project (3GPP)},
  type        = {TSG RAN Meeting \#109, TDoc RP-250766},
  year        = {2025},
  month       = {September}
}

@INPROCEEDINGS{wang_CC,
  author={Wang, Lin and others},
  booktitle={in Proc. 21st Int. Symp. Modeling Optim. Mobile, Ad Hoc, Wireless Netw. (WiOpt)}, 
  title={Coherent Compensation Based {ISAC} Signal Processing for Long-Range Sensing: (Invited Paper)}, 
  year={2023},
  doi={10.23919/WiOpt58741.2023.10349853}
}

@INPROCEEDINGS{geiger_MTCC,
  author={Geiger, Benedikt and others},
  booktitle={in Proc. 14th Int. ITG Conf. Syst., Commun. Coding (SCC)}, 
  title={Integrated Long-range Sensing and Communications in Multi Target Scenarios using {CP-OFDM}}, 
  year={2025},
  doi={10.1109/IEEECONF62907.2025.10949099}
}

@misc{Li_SIC,
      title={{OFDM-ISAC} Beyond {CP} Limit: Performance Analysis and Mitigation Algorithms}, 
      author={Li, Peishi and others},
      year={2025},
      howpublished={preprint, available at \url{https://arxiv.org/abs/2511.17878}}
}

@article{bluestein1970,
  author    = {Bluestein, Leo I.},
  title     = {A linear filtering approach to the computation of discrete {Fourier} transform},
  journal={IEEE Trans. Audio Electroacoust.},
  volume    = {18},
  number    = {4},
  pages     = {451--455},
  year      = {1970},
  publisher = {IEEE}
}

@ARTICLE{wild,
  author={Wild, Thorsten and Braun, Volker and Viswanathan, Harish},
  journal={IEEE Access}, 
  title={Joint Design of Communication and Sensing for Beyond {5G} and {6G} Systems}, 
  year={2021},
  volume={9},
  pages={30845-30857},
  doi={10.1109/ACCESS.2021.3059488}
}

@article{dinesh2013,
  author = {Bharadia, Dinesh and McMilin, Emily and Katti, Sachin},
  title = {Full duplex radios},
  year = {2013},
  month = {aug},
  volume = {43},
  number = {4},
  journal = {SIGCOMM Comput. Commun. Rev.},
  pages = {375–386},
  doi = {10.1145/2534169.2486033}
}

@ARTICLE{chafii,
  author={Chafii, Marwa and others},
  journal={IEEE Commun. Surveys Tuts.}, 
  title={Twelve Scientific Challenges for {6G}: Rethinking the Foundations of Communications Theory}, 
  year={2023},
  volume={25},
  number={2},
  pages={868-904},
  doi={10.1109/COMST.2023.3243918}
}

@INPROCEEDINGS{Nuss2018,
  author={Nuss, Benjamin and Mayer, Jonathan and Zwick, Thomas},
  booktitle={Proc. IEEE MTT-S Int. Conf. Microw. Intell. Mobility (ICMIM)}, 
  title={Limitations of {MIMO} and Multi-User Access for {OFDM} Radar in Automotive Applications}, 
  year={2018},
  doi={10.1109/ICMIM.2018.8443533}
}

@INPROCEEDINGS{mandelli_ISAC,
  author={S. Mandelli and others},
  booktitle={Proc. 6GNet}, 
  title={Survey on Integrated Sensing and Communication Performance Modeling and Use Cases Feasibility}, 
  year={2023},
  doi={10.1109/6GNet58894.2023.10317691}
}

@misc{Xu_SW,
      title={How Does {CP} Length Affect the Sensing Range for {OFDM-ISAC}?}, 
      author={Xiaoli Xu and Zhiwen Zhou and Yong Zeng},
      year={2025},
      howpublished={preprint, available at \url{https://arxiv.org/abs/2503.08062}}
}

@INPROCEEDINGS{canan_aydogdu,
  author={Aydogdu, Canan and Keskin, Musa Furkan and Wymeersch, Henk},
  booktitle={in Proc. 17th Eur. Radar Conf. (EuRAD)}, 
  title={Automotive Radar Interference Mitigation via Multi-Hop Cooperative Radar Communications}, 
  year={2021},
  doi={10.1109/EuRAD48048.2021.00076}
}

@phdthesis{Braun_diss,
  title={{OFDM} Radar Algorithms in Mobile Communication Networks},
  author={Braun, Klaus Martin},
  year={2014},
  school={Karlsruher Institut f{\"u}r Technologie (KIT)},
  doi={10.5445/IR/1000038892},
  type={Dissertation}
}

@ARTICLE{CZT_rabiner,
  author={Rabiner, L. and Schafer, R. and Rader, C.},
  journal={IEEE Trans. Audio Electroacoust.}, 
  title={The chirp {Z}-transform algorithm}, 
  year={1969},
  volume={17},
  number={2},
  pages={86-92},
  doi={10.1109/TAU.1969.1162034}
}

@misc{keskin2025,
      title={Bridging the Gap via Data-Aided Sensing: Can Bistatic {ISAC} Converge to Genie Performance?}, 
      author={Keskin, Musa Furkan and others},
      year={2025},
      howpublished={preprint, available at \url{https://arxiv.org/abs/2505.01280}}
}

@ARTICLE{lucas_overview_schemes,
  author={Giroto de Oliveira, Lucas and others},
  journal={IEEE Trans. Microw. Theory Techn.}, 
  title={Joint Radar-Communication Systems: Modulation Schemes and System Design}, 
  year={2022},
  volume={70},
  number={3},
  pages={1521-1551},
  doi={10.1109/TMTT.2021.3126887}
}

@article{Sorensen_PrunedFFT,
  author={Sorensen, H. and Burrus, C.},
  journal={IEEE Trans. Signal Process.}, 
  title={Efficient computation of the {DFT} with only a subset of input or output points}, 
  year={1993},
  volume={41},
  number={3},
  pages={1184-1200},
  doi={10.1109/78.205724}
}

@article{Richard_trust_region,
  author = {Byrd, Richard H. and Schnabel, Robert B. and Shultz, Gerald A.},
  title = {Approximate solution of the trust region problem by minimization over two-dimensional subspaces},
  year = {1988},
  month = {jan},
  journal = {Math. Program.},
  volume = {40},
  number = {1--3},
  pages = {247--263},
  doi = {10.1007/BF01580735}
}

@article{Lagarias_Nelder_Mead,
  author = {Lagarias, Jeffrey C. and others},
  title = {Convergence Properties of the Nelder--Mead Simplex Method in Low Dimensions},
  journal = {SIAM J. Optim.},
  volume = {9},
  number = {1},
  pages = {112-147},
  year = {1998},
  doi = {10.1137/S1052623496303470}
}

@ARTICLE{Su2025,
  author={Su, Yanpeng and others},
  journal={IEEE Trans. Veh. Technol.}, 
  title={{ISAC}-Enabled {OFDM} Radar: {ICI} and Off-Grid Effect Suppression via Signal Reconstruction}, 
  year={2025},
  doi={10.1109/TVT.2025.3606218}
}

@ARTICLE{Zu_ISAC_challenges,
  author={S. Lu and others},
  journal={IEEE Internet Things J.}, 
  title={Integrated Sensing and Communications: Recent Advances and Ten Open Challenges}, 
  year={2024},
  volume={11},
  number={11},
  pages={19094-19120},
  doi={10.1109/JIOT.2024.3361173}
}

@book{widrow_quantization,
  author    = {B. Widrow and I. Koll\'{a}r},
  title     = {Quantization Noise: Roundoff Error in Digital Computation, Signal Processing, Control, and Communications},
  publisher = {Cambridge Univ. Press},
  year      = {2008},
  address   = {Cambridge, U.K.}
}

@INPROCEEDINGS{karle2024modular,	
	author={C. Karle and others},
	booktitle={Proc. Int. Syst.-on-Chip Conf. (SOCC)}, 	
	title={Modular Hardware Design for High-Performance {MIMO}-Capable {SDR} Systems to Accelerate {6G} Development}, 
	year={2024},
	month={Sept.},
	doi={10.1109/SOCC62300.2024.10737765}
}

\end{document}